\documentclass{article}

\usepackage{arxiv}

\usepackage[utf8]{inputenc} 
\usepackage[T1]{fontenc}    
\usepackage{hyperref}       
\usepackage{url}            
\usepackage{booktabs}       
\usepackage{amsfonts}       
\usepackage{nicefrac}       
\usepackage{microtype}      
\usepackage{lipsum}

\usepackage{epsfig}
\usepackage[displaymath]{lineno}
\usepackage{hyperref}
\usepackage{graphicx,pstricks}
\usepackage{graphics}
\usepackage{amssymb}
\usepackage{amsmath, amsthm}
\usepackage{amsfonts}
\usepackage{algorithm}
\usepackage{booktabs,tabu,multirow}
\usepackage{subfig}
\usepackage{caption}
\usepackage{colortbl}
\usepackage{setspace}

\title{A Computational Framework for Atrioventricular Valve Modeling using Open-Source Software}

\author{
Wensi Wu\\
Department of Anesthesiology and\\
Critical Care Medicine,\\
Children's Hospital of Philadelphia,\\
Philadelphia, PA 19104 \\
\texttt{wuw4@chop.edu} \\
\And
\textbf{Stephen Ching}\\
Department of Anesthesiology and\\
Critical Care Medicine,\\
Children's Hospital of Philadelphia,\\
Philadelphia, PA 19104 \\
\texttt{chings@chop.edu} \\
\And
\textbf{Steve A. Maas}\\
Department of Biomedical Engineering, and\\
Scientific Computing and Imaging Institute, \\
University of Utah,\\
Salt Lake City, UT 84112\\
\texttt{steve.maas@utah.edu} \\  
\And
\textbf{Andras Lasso}\\
Laboratory for Percutaneous Surgery,\\
Queen’s University,\\
Kingston, ON\\
\texttt{lasso@queensu.ca} \\  
\And
\textbf{Patricia Sabin}\\
Department of Anesthesiology and\\
Critical Care Medicine,\\
Children's Hospital of Philadelphia,\\
Philadelphia, PA 19104\\
\texttt{sabinp@chop.edu} \\
\And
\textbf{Jeffrey A. Weiss}\\
Department of Biomedical Engineering, and\\
Scientific Computing and Imaging Institute, \\
University of Utah,\\
Salt Lake City, UT 84112\\
\texttt{jeff.weiss@utah.edu} \\  
\And
\textbf{Matthew A. Jolley}\\
Department of Anesthesiology and\\ 
Critical Care Medicine,\\
Division of Pediatric Cardiology,\\
Children's Hospital of Philadelphia,\\
Philadelphia, PA 19104\\
\texttt{jolleym@chop.edu} \\  
}

\begin{document}
\maketitle

\begin{abstract}
Atrioventricular valve regurgitation is a significant cause of morbidity and mortality in patients with acquired and congenital cardiac valve disease. Image-derived computational modeling of atrioventricular valves has advanced substantially over the last decade and holds particular promise to inform valve repair in small and heterogeneous populations which are less likely to be optimized through empiric clinical application. While an abundance of computational biomechanics studies have investigated mitral and tricuspid valve disease in adults, few studies have investigated application to vulnerable pediatric and congenital heart populations. Further, to date, investigators have primarily relied upon a series of commercial applications that are neither designed for image-derived modeling of cardiac valves, nor freely available to facilitate transparent and reproducible valve science. To address this deficiency, we aimed to build an open-source computational framework for the image-derived biomechanical analysis of atrioventricular valves. In the present work, we integrated an open-source valve modeling platform, SlicerHeart, and an open-source biomechanics finite element modeling software, FEBio, to facilitate image-derived atrioventricular valve model creation and finite element analysis. We present a detailed verification and sensitivity analysis to demonstrate the fidelity of this modeling in application to 3D echocardiography-derived pediatric mitral and tricuspid valve models. Our analyses achieved excellent agreement with those reported in the literature. As such, this evolving computational framework offers a promising initial foundation for future development and investigation of valve mechanics, in particular collaborative efforts targeting the development of improved repairs for children with congenital heart disease.
\end{abstract}



\section{Introduction}

Multi-modality imaging, including 3D echocardiography(3DE), has transformed adult mitral valve surgery by capturing the full, complex geometry of the valve in real time, providing an intuitive view of the functioning valve directly to the surgeon.  While informative, 3D visualization alone is insufficient for quantitative assessment and analysis of the valve. The development of image-derived mitral valve computer modeling tools has partially unlocked this potential, allowing precise, quantitative comparison of normal valves to dysfunctional valves, greatly improving the understanding of the 3D structural correlates of adult mitral valve dysfunction \cite{Salgo2002, Grewal2010, Levack2012, Lee2013}. However, correlation cannot infer causality: it does not elucidate the physical basis for why valves fail over time or allow the testing and comparison of novel repair strategies. The application of image-derived finite element modeling has begun to provide this capability, bringing forth the potential to determine patient-specific structural contributors to valve stress, strain, and failure, as well as the development of the optimal repair for an individual patient \cite{Kong2018, Villard2018, Biffi2019, Khalighi2019, Sacks2019, Kong2020}. Specifically, computational modeling allows for investigations of valve function and assessment of leaflet stress and strain that would otherwise be unobtainable using conventional in vitro methods or clinical trials \cite{Sacks2019}.

Significant advances in FEM modeling of the mitral and tricuspid valve have occurred over the last decade \cite{Lee2014, Khalighi2017, Kamensky2018, Kong2018, Sacks2019, Gryzbon2019}. Notably, several barriers to the application of FEM to the modeling of atrioventricular valves from human images have been significantly addressed through these studies. Initial FEM models were constructed from micro-CT images of ex-vivo animal and cadaveric human hearts. These high-resolution images clearly demonstrated the papillary and chordal support structure for the valve and provided a detailed roadmap for incorporation of those structures into the valve model. However, while valve leaflets and papillary muscle heads are clearly visible in living human 3DE and tomographic imaging, the individual chords are not reliably visualized. Thankfully, Khalighi et. al. compared simplified distributed chordal models to “ground truth” models based on micro-CT and showed that for chord density greater than 15 chords/cm$^2$, the results were functionally equivalent to the ground truth models \cite{Khalighi2019}. This study opened the potential to utilize living human clinical images as the basis for FEM to investigate valve leaflet stress and strain. A second advance has been the development of realistic constitutive models and specialized FEM techniques to support the requirements necessary for the modeling of valve biomechanics \cite{Drach2018, Kamensky2018, Sacks2019}. Finally, the complexity associated with modeling thin pliable structures and obtaining stable solutions (convergence) when modeling complex valve leaflet motion has been challenging \cite{Lee2014, Lee2015}. Kamensky et al. recently described a potential-based contact algorithm that is particularly well suited for valve leaflet modeling and has been subsequently validated for modeling the tricuspid valve \cite{Kamensky2018, Lee2019, Johnson2021}. The combination of these advances brings the field to a state where FEM techniques may be meaningfully applied to derived from 3D images of atrioventricular valves in living humans.

Building on this capability, previous investigations have demonstrated that perturbations of valve structure result in differences in leaflet stress and strain in image-derived models of the adult mitral valve, and more recently, the tricuspid valve \cite{Lee2014, Khalighi2017, Kamensky2018, Sacks2019, Gryzbon2019}. Notably, elevated leaflet stress and strain are associated with valve failure, pathologic changes in valve leaflets (leaflet prolapse, chordal rupture), and changes in valve leaflet gene expression \cite{Lee2013, Kodigepalli2020, Kruithof2020, Markby2020}. Further, finite element method (FEM) investigations in the mitral and tricuspid valve have demonstrated how both image derived and parametric valve models can be “surgically” altered to precisely investigate the effect of anatomic variation and surgical interventions upon leaflet stress, strain, and coaptation \cite {Lee2015, Zekry2016, Drach2018, Sacks2019, Kong2018}. These modeling capabilities may be particularly well suited to the optimization of valve repair techniques in small and heterogeneous populations, such as children with congenital heart disease, who do not benefit from empiric validation of valve repair techniques through high-volume application.

However, the majority of methods described to date leverage serial application of multiple different commercial software platforms that are not fully modifiable or configurable by the end user. A different commercial tool is used to import 3D data, create models, incorporate models in FEM software, and run simulations. Further, each commercial tool is not open or transparent as to its methodology, limiting reproducibility and extension of functionality. Similarly, suitable contact and constitutive models for valve modeling are not typically provided in such packages. As such, there is the need for the development of an open-source pipeline for computational valve modeling to catalyze open and transparent valve science. In a major step toward an integrated image-derived valve modeling pipeline, we implemented a robust constitutive model and a novel contact potential formulation in the FEBio finite element software \footnote{www.febio.org}, \cite{Maas2012, Maas2017, Ateshian2018} and applied it to 3DE image-derived models of pediatric mitral and tricuspid valves. We then performed a detailed sensitivity analysis of the effect of varying modeling parameters upon leaflet stress and strain to verify the newly implemented framework. 

\section{Methods}

In the present work, we integrated and employed open-source platforms to facilitate the computational modeling process: SlicerHeart \footnote{github.com/SlicerHeart} \cite{Scanlan2018, Nguyen2019}, 3D Slicer \footnote{www.slicer.org} \cite{Fedorov2012}, and FEBio 3.5.1 \cite{Maas2012, Maas2017, Ateshian2018}. The current pipeline consisted of two major procedures. First, we constructed 3DE-derived FEM models of atrioventricular valves using SlicerHeart and 3D Slicer. Second, we imported those valve models and performed FEM analyses to assess the mechanical responses of the valves using FEBio. Notably, while we utilized 3DE in this study, this workflow is fundamentally applicable to images created using  computed tomography (CT) or cardiac magnetic resonance imaging (MRI).

\subsection{Data Import and Valve Model Creation}

This study was approved by the Institutional review board at the Children's Hospital of Philadelphia. 3DE of mitral and tricuspid valves were identified from an existing database at the Children's Hospital of Philadelphia. The mitral valve was based on a 15 year-old male with normal heart anatomy and no mitral regurgitation. The tricuspid valve was based on a 3 year-old male with hypoplastic left heart syndrome with no significant valve regurgitation. Images had been acquired on a Philips Epiq system (Philips Medical, Andover, MA). The 3DE volume data (DICOM) were imported into 3D Slicer using the Philips 4D US DICOM patcher module in SlicerHeart, as previously described \cite {Nguyen2019}. 

\begin{figure}[h!]
\centering
\includegraphics[width=0.8\textwidth]{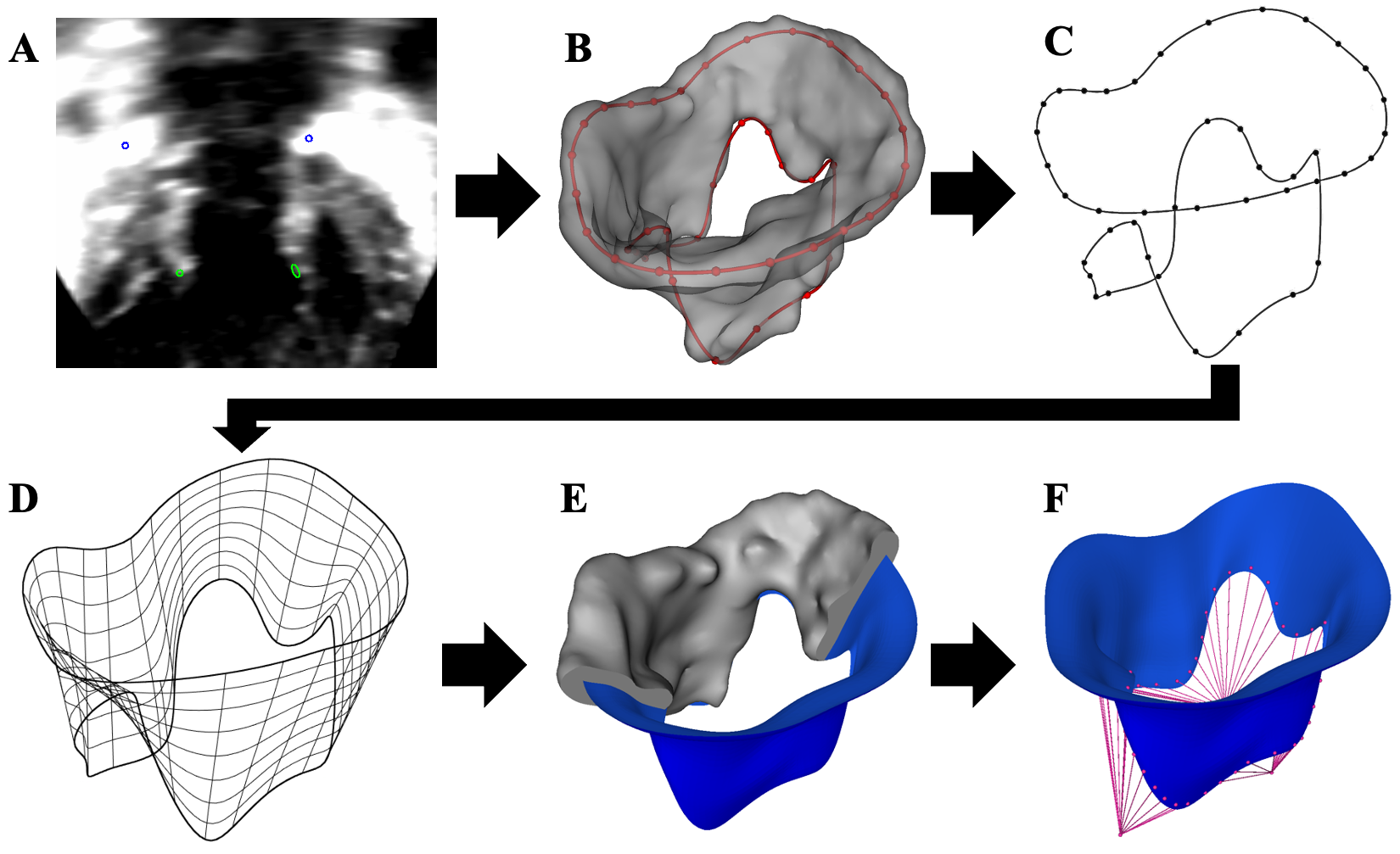}
\caption{Finite element modeling (FEM) pipeline from 3DE image: valve construction process from 3DE image to FEM model. (A) Define annulus and free edge control points, (B) create leaflet segmentation, (C) create splines, (D) initialize Non-Uniform Rational B-Spline (NURBS) surface, (E) adjust NURBS surface to segmentation, and (F) project chordae onto leaflets.}\label{pipeline}
\end{figure}

The valve model creation pipeline is described in Fig. \ref{pipeline}. First, the Valve Annular Modeling module in SlicerHeart was used to define the annulus and free edge of each valve. Then, the mitral and tricuspid valves were segmented to create valve models using the Valve Segmentation Module in SlicerHeart \cite{Nguyen2019}. The finite element valve models were created by first defining the annular contour curve using a 24-control point periodic spline. A second periodic control point spline was constructed along the free edges of the leaflets to model the open area of the valve. The resulting models were imported into Autodesk Fusion 360 (AutoDesk, San Rafael, CA). A Non-Uniform rational B-Spline (NURBS) surface was then lofted between the two curves to create the leaflets and valve surface. Control points on the NURBS surface were further edited to match the native geometry of the segmentation of the valve. This was the only part of the workflow that is not currently open-source, and the development of leaflet medial surface extraction and NURBS editing is underway in SlicerHeart to eliminate this need.

\begin{figure}[h!]
\centering
\includegraphics[width=0.8\textwidth]{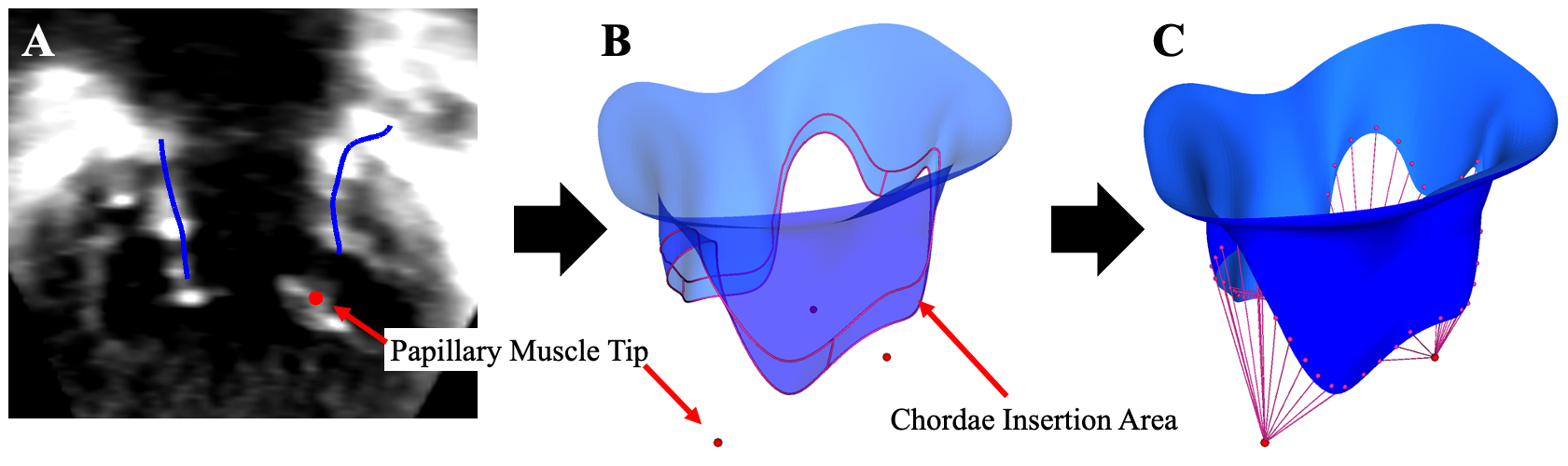}
\caption{Finite element modeling pipeline from 3D echocardiographic image: chordae tendineae modeling. (A) Identify papillary muscles, (B) define chordal insertion area, and (C) project chordae.}\label{chords}
\end{figure}

A custom Python scripted workflow was preliminarily created in SlicerHeart to distribute the chordae evenly along the leaflets (Fig. \ref{chords}). The chordae tendineae are not reliably identifiable from 3DE images, but the papillary muscle origins can readily be seen. The locations of the papillary muscle tips were identified in the 3DE image within 3D Slicer, and the registered coordinates were used to model the chordal origins as a single point per papillary muscle, as demonstrated by Khaligi et al \cite{Khalighi2017}.

\subsection{Constitutive Model of Atrioventricular Valve Leaflets} 
Atrioventricular valve leaflets consist of three layers of tissue: fibrosa, spongiosa, and atrialis. The fibrosa is the thickest layer of the three, and is directed towards the ventricular chamber, while the atrialis layer, as the name suggests, is directed towards the atria. The fibrosa layer is composed primarily of type I collagen fibers and provides the most mechanical support when subjected to flexural tension; the spongiosa layer is made of highly hydrated glycosaminoglycans (GAGs) and serves as a lubrication layer for the fibrosa and atrialis layers; the atrialis layer contains collagen and radially aligned elastin fibers and sets out to provide support for reducing radial strains when the valve undergoes physiological loading \cite{Sacks2009, maxfield2014, Sacks2019}.

In the present work we adopted an incompressible, isotropic, and hyperelastic constitutive model for the atrioventricular valve leaflet tissue, namely the Lee-Sacks constitutive model \cite{Lee2014}. While numerous studies have shown that the heart valve tissues exhibit anisotropic behaviors (\textit{i.e.,} the heart valve leaflets experience higher principal strain in the radial direction than in the circumferential direction)\cite{Stevanella2010, Wang2013, Lee2014, Lee2015, Sacks2019}, obtaining collagen fiber orientation data to accurately reflect the anisotropic characteristic feature of the leaflet tissue is a challenging task. Additionally, Wu et al. demonstrated that the effects of anisotropy on heart valve leaflet coaptation and orifice are small compared to the sizes of the leaflets \cite{Wu2018, Johnson2021}. The Lee-Sacks isotropic material model \cite{Lee2014} provided a simple and computationally efficient formulation \cite{Wu2018} and was used for approximating the biomechanic response of the leaflet tissue in the present work. The second Piola-Kirchhoff stress, $\mathbf{S}$, of the Lee-Sacks soft tissue model was formulated using an isotropic Fung-type material, wherein the contributions of the extracellular matrix and collagen fiber network were represented additively by neo-Hookean and exponential terms, respectively, to capture the nonlinear stress-strain response:

\begin{equation} \label{constitutive_model}
\mathbf{S} = 2\frac{\delta \psi_{el}}{\delta \mathbf{C}}-\lambda_p\mathbf{C}^{-1}. \\
\end{equation}
Here, $\mathbf{C}$ is the right Cauchy-Green deformation tensor, $\lambda_p$ is a Lagrange multiplier for ensuring material incompressibility ($\lambda_p$= 5000 is applied in the present work), and $\psi_{el}$ is an elastic strain energy function. The hyperelastic strain energy function was expressed as:
\begin{equation} \label{strain_function}
\psi_{el}= \frac{c_0}{2}(I_1-3)+\frac{c_1}{2}(\exp{c_2(I_1-3)^2}),
\end{equation}
where $c_0$, $c_1$, and $c_2$ are material coefficients, and $I_1 = tr \mathbf{C}$. Therefore, the term $\frac{\delta \psi_{el}}{\delta \mathbf{C}}$ in Eq. \ref{constitutive_model} becomes: 
\begin{equation} \label{strain_function_diff}
\frac{\delta \psi_{el}}{\delta \mathbf{C}} = \frac{1}{2}(c_0+2c_1c_2(I_1-3)\exp{c_2(I_1-3)^2})\mathbf{I}.
\end{equation}

\subsection{Chordae Tendineae Modeling}
The chordae tendineae play a critical role in ensuring proper heart valve closure; these branch-like collagenous tissues connect the leaflets to the papillary muscle heads to prevent leaflet prolapse into the left atrium during ventricular contraction. However, while individual chordae can be visualized via \textit{ex-vivo} CT scan, individual chordae cannot be be reliably visualized using clinically derived 3DE in humans\cite{Sacks2019}. Therefore, the geometric modeling of chordae tendineae from 3DE derived models relies on strategic simplifications to circumvent modeling limitations while accurately preserving their biological function. In the present work, we opted for a simplified, but robust, chordal topology approach introduced by Khalighi et al \cite{Khalighi2017}. Khalighi et al demonstrated that a uniformly distributed, branchless chordal model with $15\pm2$ chords/cm$^2$ was able to reproduce the ground truth results in predicting mitral valve closure to a high degree of accuracy. We uniformly distributed at least 17 chords/cm$^2$ over the free edge of our valve models. The chords were modeled as tension-only 2-node linear springs that connect the leaflet insertion points to the papillary muscle tips. The mechanical behavior of the springs was considered mathematically within FEBio with a nonlinear force-displacement response where the tension was equal to zero up to a defined stretch value and the springs behaved linearly once the stretch threshold was reached. 

\subsection{Contact Potential}
Atrioventricular valve leaflet contact can be particularly difficult to realistically model. Kamensky et al. recently described a potential-based contact formulation \cite{Kamensky2018}. This contact formulation was implemented in FEBio after preliminary investigation found it to be particularly well suited to allowing the valve leaflets to slide past one another, avoiding unrealistic penetration of opposed leaflets. The contact potential between the two objects of interest was approximated by the contact potential energy expressed as:

\begin{equation} \label{contact_potential}
E_c = \int_{\Omega_0^2}  \int_{\Omega_0^1} \phi(r_{12}) \,d\mathbf{X}_1 d\mathbf{X}_2 ,
\end{equation}
where $\mathbf{X}_1$ and $\mathbf{X}_2$ are in the reference configurations of the two contact bodies denoted by $\Omega_0^1$ and $\Omega_0^2$ respectively, $\phi(r_{12})$ is a contact potential kernel, and $r_{12}$ denotes the Euclidean distance between two contact points.

The contact potential was described by the following force-separation law:
\begin{equation} \label{force_seperation_law}
\begin{gathered}
- \phi^{'}(r_{12}) = 
\begin{cases}
   \frac{k_c}{(r_{12})^p}-c_2,   & r_2 < r_{in}, \\
    c_1(r_{12}-r_{out})^2,  & r_{in} \leq r_{12} < r_{out}, \\
    0, &\text{otherwise},
\end{cases}
\end{gathered}
\end{equation}
where $k_c$ is a dimensionless scaling factor for the contact force, $r_{in}$ is the inner distance that governs the transition between parabolic and higher order regions of the contact potential, $r_{out}$ is the outer distance that defines the boundary of the contact surface, and $p$ is the power of the contact potential. We refer interested readers to \cite{Kamensky2018, Johnson2021} for detailed information regarding the numerical implementation of the contact potential. 

\subsection{Solution Procedure}
The valve medial surface was discretized into 4-node linear quadrilateral (Quad4) shell elements \cite{Hou2018}. We applied fixed displacement boundary conditions to the outer circumference of the annulus edge and the papillary muscle tips. We deformed the model from diastolic to systolic configuration by prescribing a physiologically realistic systolic ventricular pressure orthogonal to the ventricular surface of the leaflets. In contact models, it is not unusual for the simulation to fail to converge because of unrealistic high frequency modes in the numerical solution. We employed a mass-damping scheme by applying a damping matrix $C = 2000$~Ns/m to the model to suppress any such spurious oscillations. The implicit Newmark time integration scheme, along with an automatic time stepping algorithm, was used to facilitate time advancement and maintain nonlinear solution convergence in the transient dynamic simulation. The displacement and energy residual tolerances in FEBio were set to 0.001 and 0.1, respectively.

\begin{figure}[h!]
\centering
\includegraphics[width=0.8\textwidth]{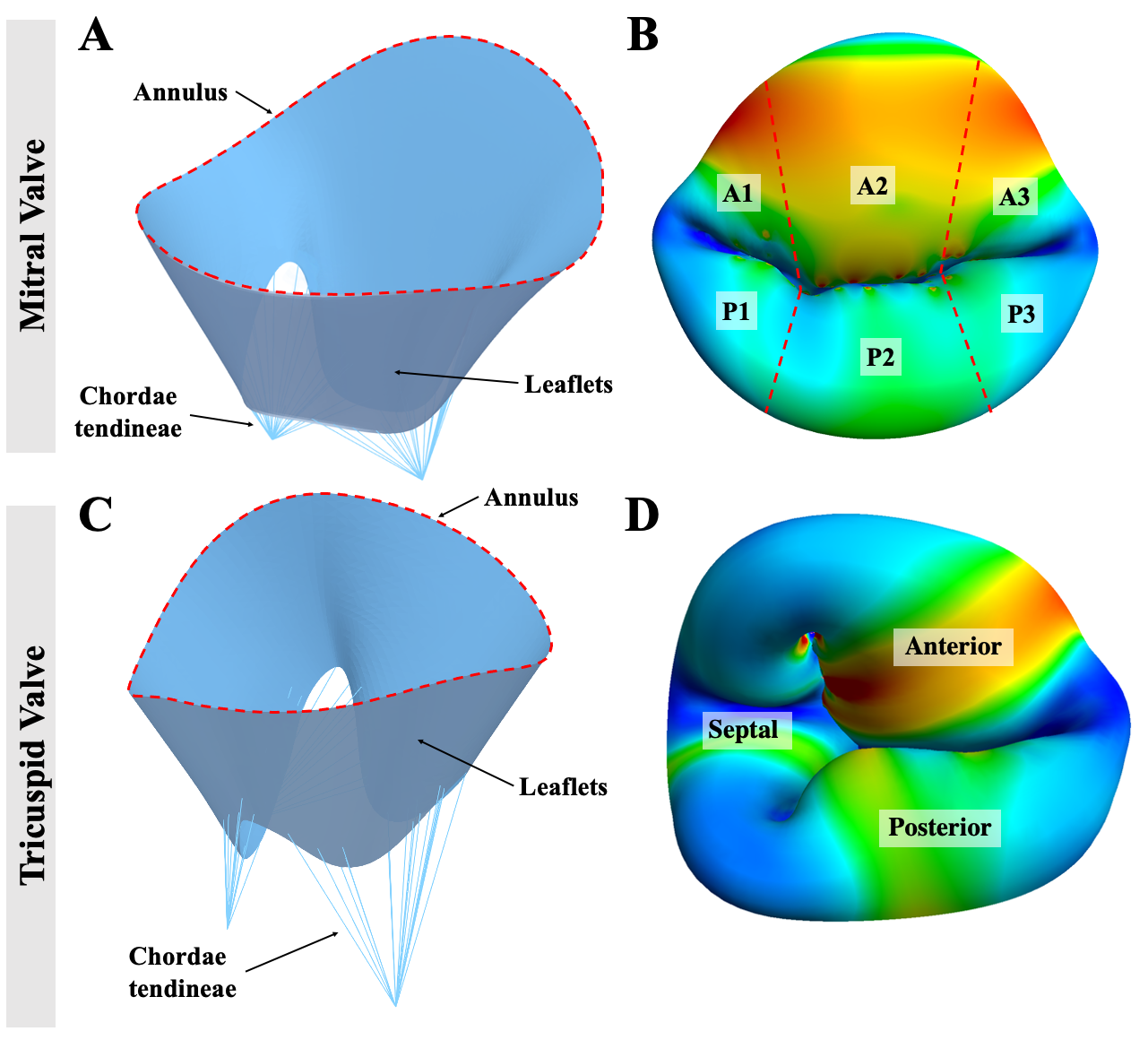}
\caption{Geometric representation of image-derived mitral and tricuspid valve finite element models. (A) Model of open mitral valve with annulus, leaflets, and chordae tendineae defined, (B) model of closed mitral valve with leaflet regions defined, (C) model of an open tricuspid valve with annulus, leaflets, and chordae tendineae defined, and (D) model of a closed tricuspid valve with leaflets defined. Mitral valve annular circumference was 12.3 cm and annular area projected onto the least squares annular plane was 11.0 cm$^2$. Tricuspid valve annular circumference was 13.1 cm and annular area projected onto the least squares annular plane was 12.0 cm$^2$.}\label{mitral_tricuspid_geo}
\end{figure}

Fig. \ref{mitral_tricuspid_geo}A and \ref{mitral_tricuspid_geo}B provide a geometric visualization of the mitral valve model. The mitral model was discretized into 735 elements, 2965 elements, and 11689 elements for mesh convergence analysis. The material coefficients used for the mitral valve were: $c_0 = 200$ kPa; $c_1  = 2968.4$ kPa; and $c_2=0.2661$ \cite{Lee2014}. We assumed 0.396 mm uniform thickness over the leaflets.

We applied 100 mm Hg ventricular pressure to the ventricular surface of the mitral valve. The pressure was first increased linearly over 0.005~s (consistent with \cite{Lee2014}) to initiate the dynamic simulations. The pressure was then kept constant until the valve reached steady-state (full closure). The pressure evolution as a function of time was: 

\begin{equation}\label{tension_force}
\begin{gathered}
p(t) = 100~\text{mm Hg}
\begin{cases}
 t/T,   & t< T~\text{s}, \\
  1,  & \text{otherwise},
\end{cases}
\end{gathered}
\end{equation}

\noindent where $T$ is the ramp-up time scale and was set to 0.005~s. Once $T$ is reached, the pressure load is maintained until steady-state. The initial time step increment was set to $\Delta t = 0.0001$~s. We terminated the simulation at 0.02~s. A total of 57 chords were attached to the mitral valve. The chordal force as a function of leaflet displacement was described by a step function in which zero tethering force was applied for the initial 5 mm stretch threshold to accurately reflect the chordal slack length. The chords were each prescribed a 30 mN tension force after the stretch threshold. The contact potential parameters for the fine mesh were: $k_c = 1$, $p = 4$, $R_{in} = 0.2$ mm, and $R_{out} = 0.5$ mm.  

A visualization of the tricuspid valve model is shown in Fig. \ref{mitral_tricuspid_geo}C and \ref{mitral_tricuspid_geo}D. The mesh densities for the coarse, medium, and fine meshes consisted of 761, 3044, and 12176 elements, respectively. The material coefficients used for the tricuspid valve were: $c_0 = 10$ kPa; $c_1  = 0.209$ kPa; and $c_2=9.046$ \cite{Stevanella2010, Kamensky2018}. We assumed 0.396 mm uniform thickness over the leaflets. The peak systolic pressure value, 23.7~ mm Hg \cite{Stevanella2010, Kong2018}, was applied to the tricupsid model: 

\begin{equation}\label{tension_force}
\begin{gathered}
p(t) = 23.7~\text{mm Hg}
\begin{cases}
 t/0.01,   & t< 0.01~\text{s}, \\
  1,  & \text{otherwise}.
\end{cases}
\end{gathered}
\end{equation}

\noindent We attached a total of 41 chords to the tricuspid valve; 20 mN chordal tension force was applied to each chord. The contact potential parameters for the fine mesh were: $k_c = 1$, $p = 4$, $R_{in} = 0.25$ mm, and $R_{out} = 0.4$ mm.

All numerical simulations were performed on a high performance computing system located at the Children's Hospital of Philadelphia, consisting of Intel Xeon CPU E5-2680 v3 computing nodes. Each node contains 24 cores, with 2.5 GHz core speed and 128 GB RAM per node. The CPU time of the mitral and tricuspid valve fine mesh models were 14 and 34 minutes, respectively.

\subsection{Convergence Analysis}
We examined the mean, 75$^\textrm{th}$ percentile, and 95$^\textrm{th}$ percentile of the $1^\textrm{st}$ principal stress and strain, together with the valve closure profiles of the three mesh models to assess the convergence. The mean values represent the average stress/strain over all elements. The 75$^\textrm{th}$ percentile and 95$^\textrm{th}$ percentile refer to the values that cover 75$\%$ and 95$\%$ of the stresses/strains, respectively. Mesh convergence was determined when the percentage of $L^2$ relative error norms reduced to less than $5\%$. The $L^2$ relative error norms were defined as: 

\begin{equation}\label{tension_force}
L^2 \text{ relative error norm}= \sqrt{\frac{\sum_{i=0}^N(S_{ref,i}-S_{test,i})^2}{\sum_{i=0}^NS_{ref,i}^2}} \times 100,
\end{equation}

\noindent where $S_{ref, i}$ represents the $i^\textrm{th}$ index of the stress or strain time history with the fine mesh, $S_{test, i}$ represents the $i^\textrm{th}$ index of the stress or strain time history with either the coarse or medium mesh, and $N$ represents the total number of time steps. 

\subsection{Uncertainty Analysis}
Uncertainty analysis is an indispensable part of computational modeling. With numerous parameters involved in valve modeling, there is a need to better understand the role each parameter plays in affecting the final model responses. As such, rigorous sensitivity analyses were performed to identify the effect of the parameters and inform the robustness and predictive capability of the numerical models and approaches. Ultimately, such analyses correlate \textit{in vivo} and \textit{in silico} data such that reasonable tolerances and margins of error can be determined should such models be used in a translational setting.

We carried out two uncertainty analysis approaches (traditional and statistical) with the fine mesh mitral and tricuspid models. In the traditional approach, we explored the influence of the modeling parameters individually, while in the statistical approach, we utilized a polynomial chaos expansion (PCE) function to quantify the uncertainty in the FEM model \cite{Burk2020}. The parameters under consideration were the material coefficients in the Lee-Sacks constitutive models (coefficient $c_0$ in the neo-Hookean term, and coefficients $c_1$ and $c_2$ in the isotropic exponential term) \cite{Lee2014}. Additionally, we explored the influence of the chordal stretch threshold and tension, as previous works suggested chordal rest length had significant impacts on valve closures \cite{becker2011, Mansi2012, Grbic2017}. We sampled five data points uniformly distributed within the range of $\pm 50\%$ of the baseline values used in the mesh convergence analyses.

We leveraged UncertainSCI for the statistical uncertainty analysis procedure, a Python-based toolkit that harnesses modern techniques to estimate model and parametric uncertainty, with a particular emphasis on needs for biomedical simulations and applications \cite{Burk2020}. This was achieved by interfacing UncertainSCI with the FEBio solver using via  a Python subroutine within FEBio and FEBioUncertainSCI \footnote{github.com/febiosoftware/FEBioUncertainSCI}. UncertainSCI used a polynomical chaos expansion function with a weighted approximate Fekete points (WAFP) method to randomly generate collocation points within the user-specified n-dimensional space (where the number of dimensions is determined by the number of input parameters) for sensitivity quantification. Unlike the traditional approach, where the parameters of interest were varied one at a time, UncertainSCI allowed for multiple variables such that the influence of input interactions may be examined. Furthermore, UncertainSCI provided quantitative measures of the uncertainties from the input parameters by computing the relative variance that each parameter contributed to the total variance, namely the first-order Sobel index. This enables better evaluation and comparison of the model output uncertainty from the material constants.

\section{Results}

We present verification and sensitivity analysis results of two image-derived valve models (mitral and tricuspid) to demonstrate the feasibility and robustness of the open-source software in the present work. For each valve model, we considered three mesh densities (coarse, medium, and fine). The length and width of each element were divided exactly in half within each level of refinement -- resulting in a fourfold increase in mesh density as we refined the models. After we established the convergence of the FEM models, we used the fine mesh models to perform uncertainty quantification to study the effect of modeling input parameters on the biomechanical responses of atrioventricular valves.  

\subsection{Mitral Valve Verification}
The means, standard deviations, 75$^\textrm{th}$ percentile, and 95$^\textrm{th}$ percentile of the stresses and strains of the whole mitral valve are shown in Fig. \ref{mitral_ver}C and \ref{mitral_ver}D. We did not see substantial differences in stresses and strains related to mesh density. The sum of chordal tethering force on both papillary muscles was 6.84~N.

Each mitral valve leaflet is anatomically divided into three scallop regions. Fig. \ref{mitral_ver}E and \ref{mitral_ver}F present the 95$^\textrm{th}$ percentile regional stresses and strains with the fine mesh. The results suggested that the anterior leaflet experiences a higher stress concentration (about two times higher than the posterior leaflet). The stress levels among the three scallop regions within each leaflet were similar. In our mitral model the region A2 experienced lower strains compared to regions A1 and A3. However, region P2 presented relatively higher strains in comparison to regions P1 and P3. 

\begin{figure}[h!]
\centering
\includegraphics[width=0.8\textwidth]{figure_updates/Figure_04.png}
\caption{Mitral valve stress and strain responses. (A) Stress profile on the mitral valve at steady state, (B) strain profile on the mitral valve at steady state, (C) stress responses with coarse, medium, and fine meshes (shaded areas indicate standard deviations), (D) strain responses with coarse, medium, and fine meshes (shaded areas indicate standard deviations), (E) 95$^\textrm{th}$ percentile 1$^\textrm{st}$ principal stress responses on various mitral valve regions, and (F) 95$^\textrm{th}$ percentile 1$^\textrm{st}$ strain responses on various mitral valve regions. Results suggested that the anterior leaflet experiences higher stress and strain concentrations than the posterior leaflet.}
\label{mitral_ver}
\end{figure}

\begin{figure}[h!]
\centering
\includegraphics[width=0.8\textwidth]{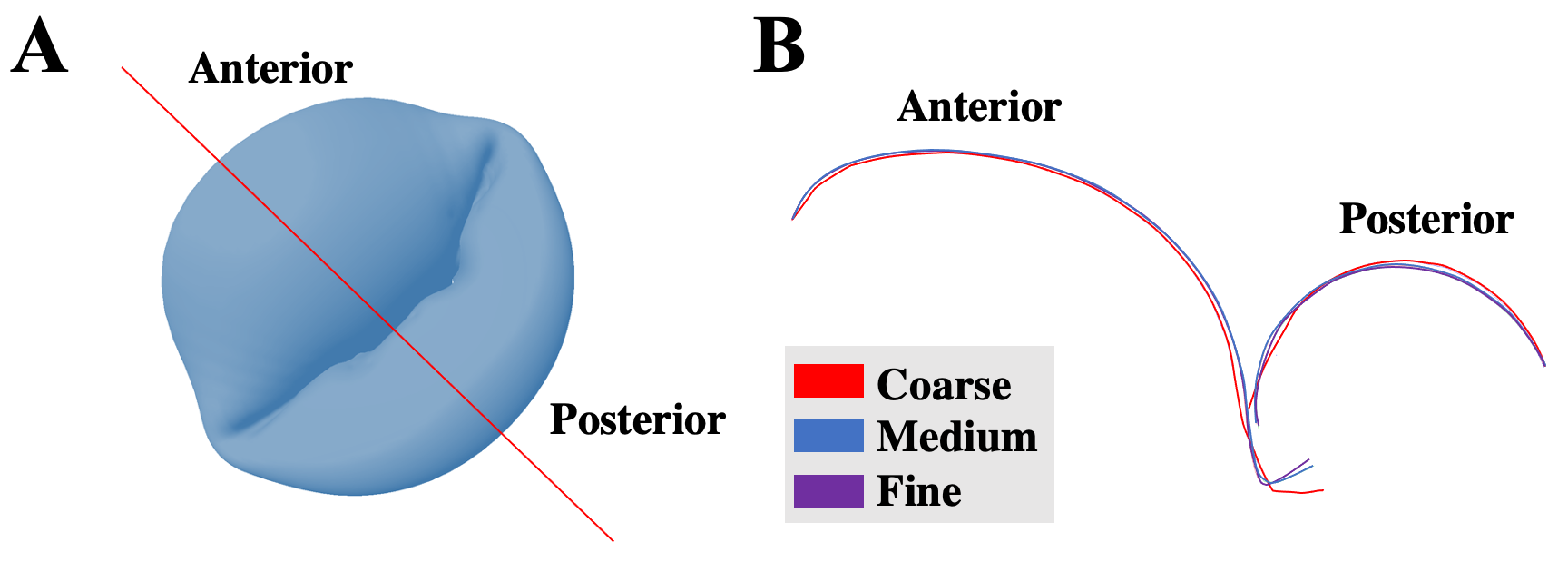}
\caption{Mitral valve closing profiles. (A) Location at which the slices were made (red line), and (B) valve closure configurations for coarse, medium, and fine meshes at the anterior-posterior coaptation. Nearly identical closing profiles suggested that all of the tested mesh densities were sufficient to capture valve closing behavior.}\label{mitral_slices}
\end{figure}

We evaluated a cross section of the mitral valve to assess valve closure of the three meshes (Fig. \ref{mitral_slices}). The closing profiles were nearly identical among the three meshes, which suggested that the mesh densities used in the present work were sufficient to capture the valve closing behavior. 

The convergence behavior was confirmed by the $L^2$ relative error norms, where we observed substantial error reduction with higher mesh density (Table \ref{error_norm}). Region P1 reported the highest percentage error in mesh refinement level from medium to fine, with $4.89\%$ relative error in mean principal stress. Nonetheless, this relative error was below the $5\%$ threshold; indicating reasonable agreement between the medium and fine meshes. The mean, $75^\textrm{th}$ percentile, and $95^\textrm{th}$ percentile stresses of the whole mitral with the fine mesh at steady state were 252.29~kPa, 336.45~kPa, and 546.22~kPa, respectively. The corresponding mean, $75^\textrm{th}$ percentile, and $95^\textrm{th}$ percentile strains at steady state were 0.21, 0.25, and 0.34, respectively. Regional stresses and strains are reported in Appendix \ref{appendix:a} Table \ref{mitral_responses_summary}.

\begin{table}[h!]
\centering
\resizebox{\textwidth}{!}{\begin{tabular}{cccccccc}
\toprule
 && \multicolumn{3}{c}{$1^\textrm{st}$ principal stress} & \multicolumn{3}{c}{$1^\textrm{st}$ principal strain} \\\cline{3-8}
 &Mesh refinement       & mean     & $75^\textrm{th}$ percentile & $95^\textrm{th}$ percentile & mean      & $75^\textrm{th}$ percentile & $95^\textrm{th}$ percentile\\
\midrule  
\multirow{2}{*}{Mitral} &coarse to medium   &  3.27   &  1.93  &  1.72   &    1.46      &     2.09  &     1.22  \\
 & medium to fine     &  0.92       &  0.74       &  0.53    &    0.39     &     0.77        &  0.81 \\
\rowcolor{lightgray}
 &coarse to medium   &  3.76   &  2.19  &  2.61  &    5.36      &     1.79   & 1.83     \\
\rowcolor{lightgray}
\multirow{-2}{*}{A1}    & medium to fine      &  0.99      &  0.90    &  0.73   &     1.66     &     0.61      &     0.50  \\
\multirow{2}{*}{A2}     &coarse to medium   &  1.98       &  1.97   &  1.18  &    0.96      &     1.19  &     0.90        \\
 & medium to fine   &  0.65    &  0.52     &  0.73   &     0.45     &     0.42 &     0.88           \\
\rowcolor{lightgray}
 &coarse to medium    &  2.07      &  1.16     &  2.10    & 1.02  &     1.05     &     1.04  \\
\rowcolor{lightgray}
\multirow{-2}{*}{A3}     & medium to fine    &  1.16     &  1.09 &    0.70   &     1.17     &     0.73                      &     0.63     \\
\multirow{2}{*}{P1}     &coarse to medium     &  12.12    &  9.21    &  10.22   &    9.80      &     10.60   &     6.69    \\
  & medium to fine           &  4.89    &  3.37  &  3.79      &     2.88     &     3.00      &     3.10                    \\
\rowcolor{lightgray} 
&coarse to medium    &  10.17 &  8.14   &  7.16   &    5.44      &     5.88   &    7.15 \\
\rowcolor{lightgray}
\multirow{-2}{*}{P2}    & medium to fine    & 2.10     &  2.05   &  2.41   &     0.92     &     1.36 &     1.70     \\
\multirow{2}{*}{P3}     &coarse to medium    &  4.03      &  5.84  &  11.22   &    5.03      &     6.74    &     1.22   \\
& medium to fine  &  1.66     &  2.11 &  2.69  &    1.29     &     1.67    & 0.81  \\
\bottomrule
\end{tabular}}
\caption{$L^2$ relative error norms (\%) in various sections of the mitral valve. All $L^2$ relative error norms were under $5\%$ in the medium to fine mesh refinement level, with the highest error norm ($4.89\%$) in the mean $1^\textrm{st}$ principal stress in region P1.} \label{error_norm} 
\end{table} 

\newpage
\subsection{Tricuspid Valve Verification}
\begin{figure}[h!]
\centering
\includegraphics[width=0.8\textwidth]{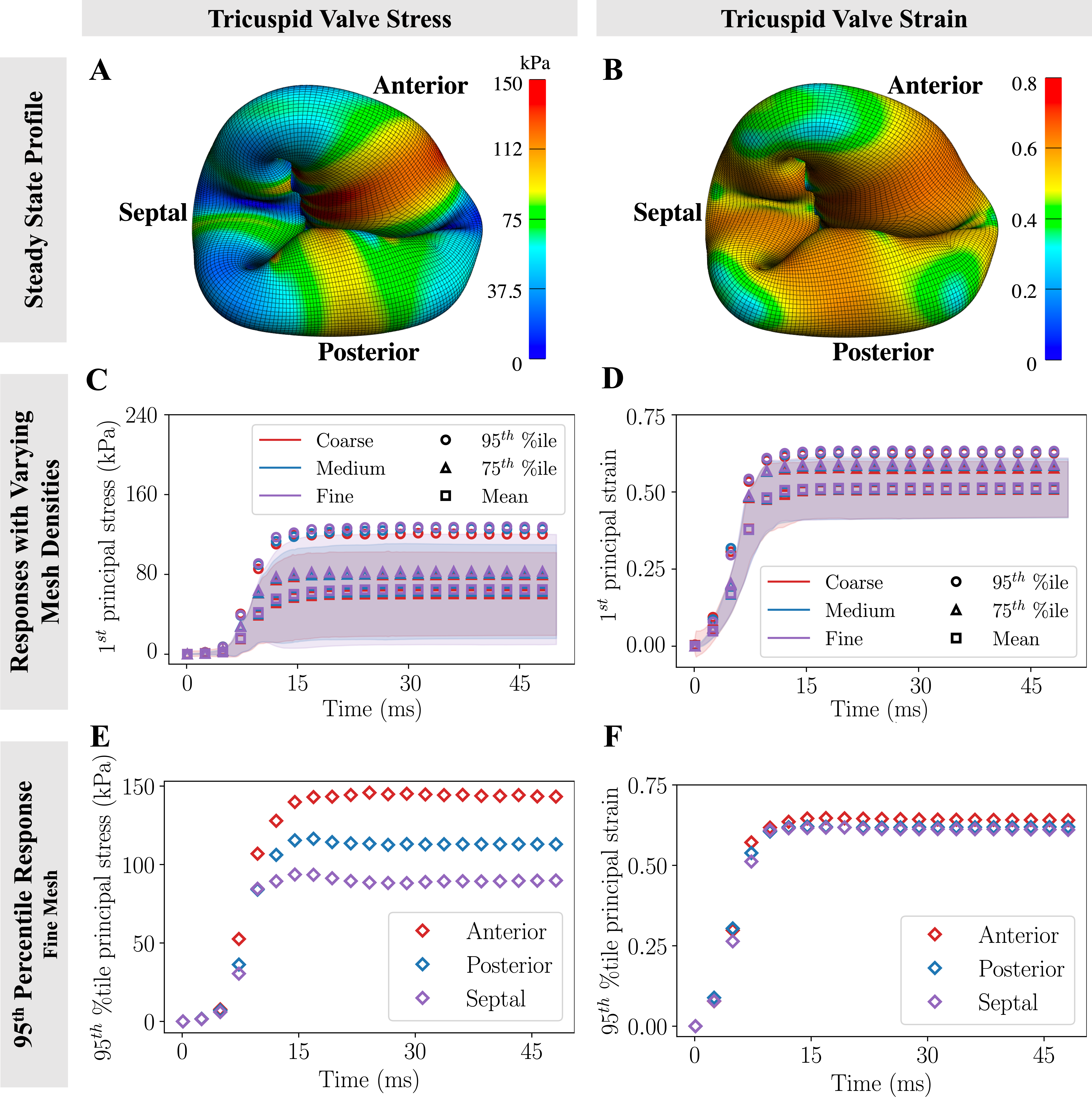}
\caption{Tricuspid valve stress and strain responses. (A) Stress profile on the tricuspid valve at steady state, (B) strain profile on the tricuspid valve at steady state, (C) stress responses with coarse, medium, and fine meshes (shaded areas indicate standard deviations), (D) strain responses with coarse, medium, and fine meshes (shaded areas indicate standard deviations), (E) 95$^\textrm{th}$ percentile 1$^\textrm{st}$ principal stress responses on various tricuspid valve leaflets, (F) 95$^\textrm{th}$ percentile 1$^\textrm{st}$ principal strain responses on various tricuspid valve regions. While results suggested that the anterior leaflet experienced higher stress concentrations than the posterior and septal leaflets, 95$^\textrm{th}$ percentile 1$^\textrm{st}$ principal strains were nearly identical among the leaflets.}
\label{tricuspid_ver}
\end{figure}

\begin{figure}[h!]
\centering
\includegraphics[width=0.8\textwidth]{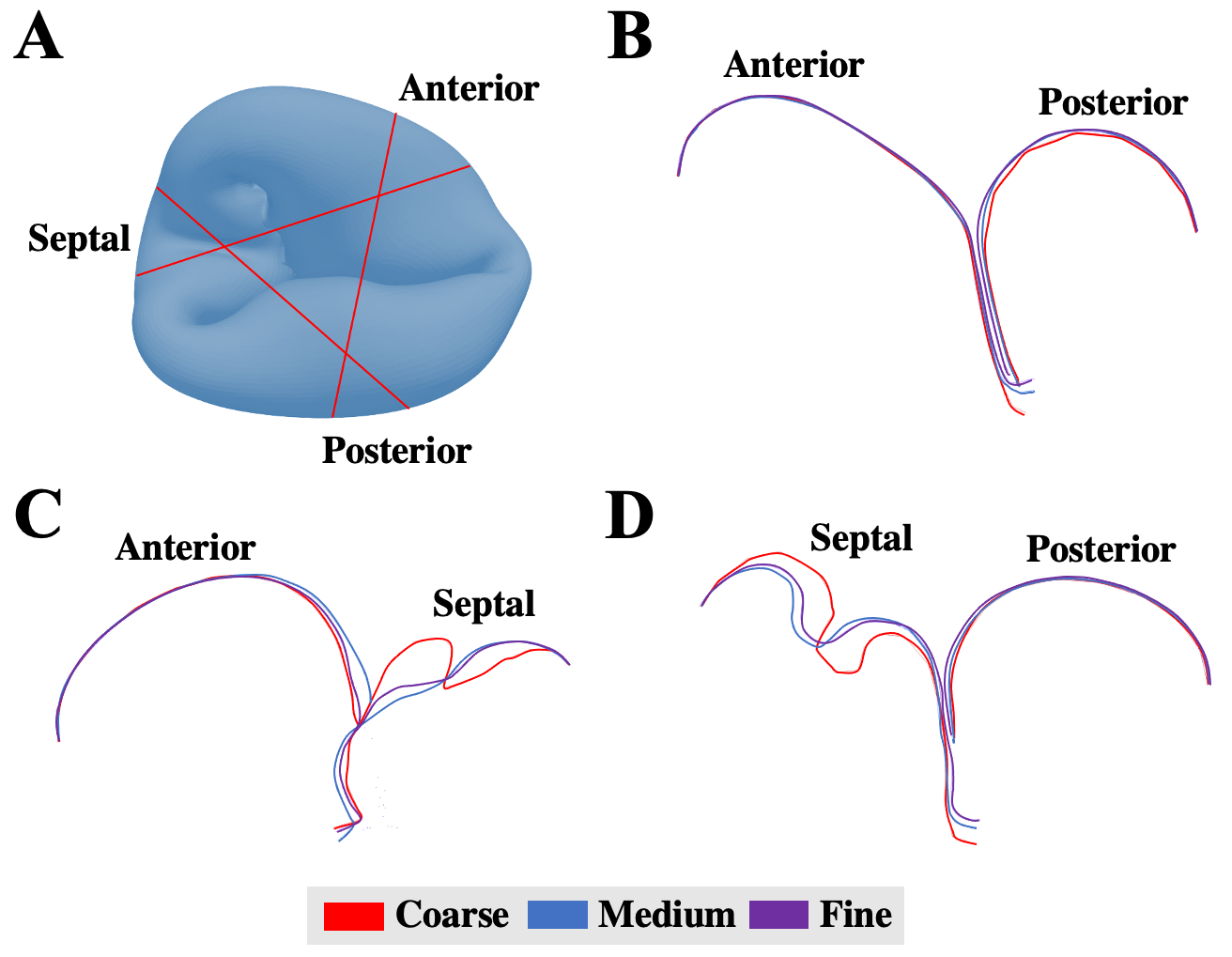}
\caption{Tricuspid valve closing profiles. (A) Locations at which the slices were made (red lines); valve closure configurations for coarse, medium, and fine meshes at the (B) anterior-posterior coaptation, (C) anterior-septal coaptation, and (D) posterior-septal coaptation. Septal leaflet deformation with the coarse mesh showed significantly different characteristics in comparison to the two finer meshes, suggesting that the mesh density played an important role in capturing complex curvatures, as demonstrated in the tricuspid model.}\label{tricuspid_slices}
\end{figure}

The stress and strain responses of the tricuspid valve are shown in Fig. \ref{tricuspid_ver}. Fig. \ref{tricuspid_ver}E suggests that the anterior leaflet experienced the highest stress concentration, followed by the posterior leaflet, and lastly the septal leaflet. On the other hand, the $95^\textrm{th}$ percentile of strains were nearly identical among the leaflets (Fig. \ref{tricuspid_ver}F). Similarly to the mitral valve, we created three cross sections through the tricuspid valve to assess the closure configurations between leaflets (Fig. \ref{tricuspid_slices}). Differences in valve closure profiles were observed with the models. The septal leaflet deformation with the coarse mesh showed significantly different characteristics in comparison to the two finer meshes. This suggested that the mesh density played an important role in capturing complex curvatures, as demonstrated in our specific image-derived tricuspid model. The sum of chordal tethering force on all papillary muscles was 2.38~N.

The $L^2$ relative error norms of the tricuspid model are presented in Table \ref{error_norm_tri}. The highest percentage error (in mesh refinement level from medium to fine) was found in the posterior leaflet, with $4.8\%$ in the $95^\textrm{th}$ percentile principal stress. All relative errors were below $5\%$. Therefore, the fine mesh of tricuspid model was a converged mesh. The mean, $75^\textrm{th}$ percentile, and $95^\textrm{th}$ percentile stresses of the whole tricuspid valve at steady state were 64.83~kPa, 82.62~kPa, and 127.27~kPa, respectively. The mean, $75^\textrm{th}$ percentile, and $95^\textrm{th}$ percentile strains at steady state were 0.51, 0.59, and 0.63, respectively; stresses and strains on individual leaflet are reported in Appendix \ref{appendix:a} Table \ref{tricuspid_responses_summary}.

\begin{table}[h!]
\centering
\resizebox{\textwidth}{!}{\begin{tabular}{cccccccc}
\toprule
 && \multicolumn{3}{c}{$1^\textrm{st}$ principal stress} & \multicolumn{3}{c}{$1^\textrm{st}$ principal strain} \\\cline{3-8}
 &Mesh refinement       & mean     & $75^\textrm{th}$ percentile & $95^\textrm{th}$ percentile & mean      & $75^\textrm{th}$ percentile & $95^\textrm{th}$ percentile\\
\midrule  
\multirow{2}{*}{Tricuspid} &coarse to medium   &  6.94  &  3.80  &  5.18   &    1.38      &     2.22                      &     1.65       \\
 & medium to fine    &  3.33   &  2.63   &  2.05   &    0.59     &    0.86   &     1.62  \\
\rowcolor{lightgray} 
 &coarse to medium  &  5.25  &  1.17  &  2.46  &    1.39   &  0.78  &     2.04  \\
\rowcolor{lightgray}
\multirow{-2}{*}{Anterior}    & medium to fine     &  4.28   &  1.49 &  4.02  &     0.80     &     0.88  &     1.85  \\
\multirow{2}{*}{Posterior}     &coarse to medium  & 5.58    &  4.80   &  9.56  &  1.10      &     2.59   &    3.77    \\ 
& medium to fine  &  3.27     &  3.12   &  4.80   &     0.56     &   1.09   &      1.52  \\
\rowcolor{lightgray}
&coarse to medium   &  10.78 &  5.73  &  3.57   &    3.33  &  3.59   &     2.92      \\
\rowcolor{lightgray}
\multirow{-2}{*}{Septal}   & medium to fine &  2.25  &  2.41  &  1.78  &  0.64  &     1.04   &     0.76\\

\bottomrule
\end{tabular}}
\caption{$L^2$ relative error norms (\%) in various sections of the tricuspid valve. All $L^2$ relative error norms were under $5\%$ in the medium to fine mesh refinement level, with the highest error norm ($4.8\%$) in the $95^\textrm{th}$ percentile of the $1^\textrm{st}$ principal stress in the posterior leaflet.} \label{error_norm_tri} 
\end{table} 

\subsection{Traditional Sensitivity Analysis}
\begin{figure}[h!]
\centering
\includegraphics[width=0.8\textwidth]{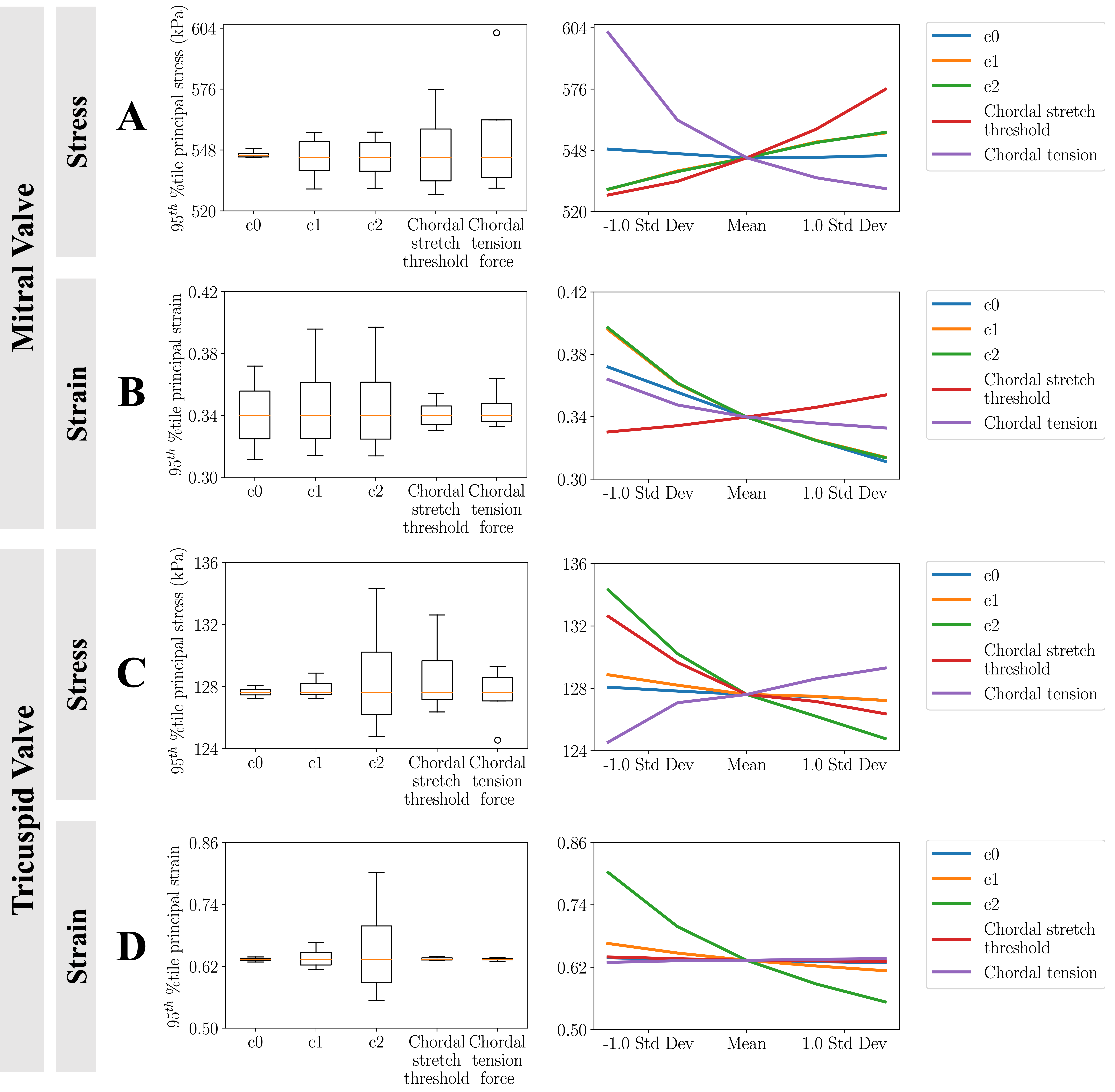}
\caption{The resulting sensitivity in the 95$^\textrm{th}$ percentile 1$^\textrm{st}$ principal stresses and strains from various modeling parameters. (A) Sensitivity of stress in the mitral valve, (B) sensitivity of strain in the mitral valve, (C) sensitivity of stress in the tricuspid valve, and (D) sensitivity of strain in the tricuspid valve. The main box-plots demonstrate the spread of skewness of the stresses and strains for each individual modeling parameter. The line plots display the relations between the modeling parameters and the stresses/strains. Results suggested that the mitral valve stresses and strains were most sensitive to material coefficients $c_1$ and $c_2$, with $c_0$ displaying additional influence on strains; tricuspid valve stresses and strains were most sensitive to material coefficient $c_2$. For the mitral valve, increased chordal stretch threshold led to higher stress/strain and increased chordal tension led to lower stress/strain, whereas the tricuspid valve displayed the opposite responses for stress and negligible responses for strain.}
\label{traditional_sensitivity}
\end{figure}

The $95^\textrm{th}$ percentile of the 1$^\textrm{st}$ principal stresses and strains at steady state are presented in Fig. \ref{traditional_sensitivity}. The mitral and tricuspid valve modeling parameters demonstrated different influences on the stress and strain responses in the valve leaflets. Among the material coefficients for the mitral valve, the material coefficients in the isotropic exponential term ($c_1$ and $c_2$) had the most significant effect on both stresses and strains -- with approximately the same minimum, maximum, and interquartile range (Fig. \ref{traditional_sensitivity}A and \ref{traditional_sensitivity}B); material coefficient $c_0$ had an inconsequential influence on the stresses, but had a strong influence on the strain response. Additionally, Fig. \ref{traditional_sensitivity}A and \ref{traditional_sensitivity}B show that the chordal stretch threshold and tension force had opposite effects in stresses and strains. We observed that a higher chordal stretch threshold lead to higher stresses and strains; contrarily, increasing chordal tension forces yielded reduced mitral valve mechanical responses. 

An opposite behavior regarding the effect of the chordal stretch threshold and tension force was seen in the tricuspid valve stress responses (Fig. \ref{traditional_sensitivity}C). Specifically, great chordal tension and lower chordal stretch threshold correspond to higher leaflet stress. Furthermore, Fig. \ref{traditional_sensitivity}D indicates the chordal stretch threshold and tension force had insignificant effects on tricuspid strains. In terms of material coefficients, Fig. \ref{traditional_sensitivity}C and \ref{traditional_sensitivity}D suggest that the tricuspid model was most sensitive to the coefficient $c_2$ (in agreement with the mitral model). In contrast, coefficients $c_0$ and $c_1$ had negligible effects. 

\begin{figure}[h!]
\centering
\includegraphics[width=0.8\textwidth]{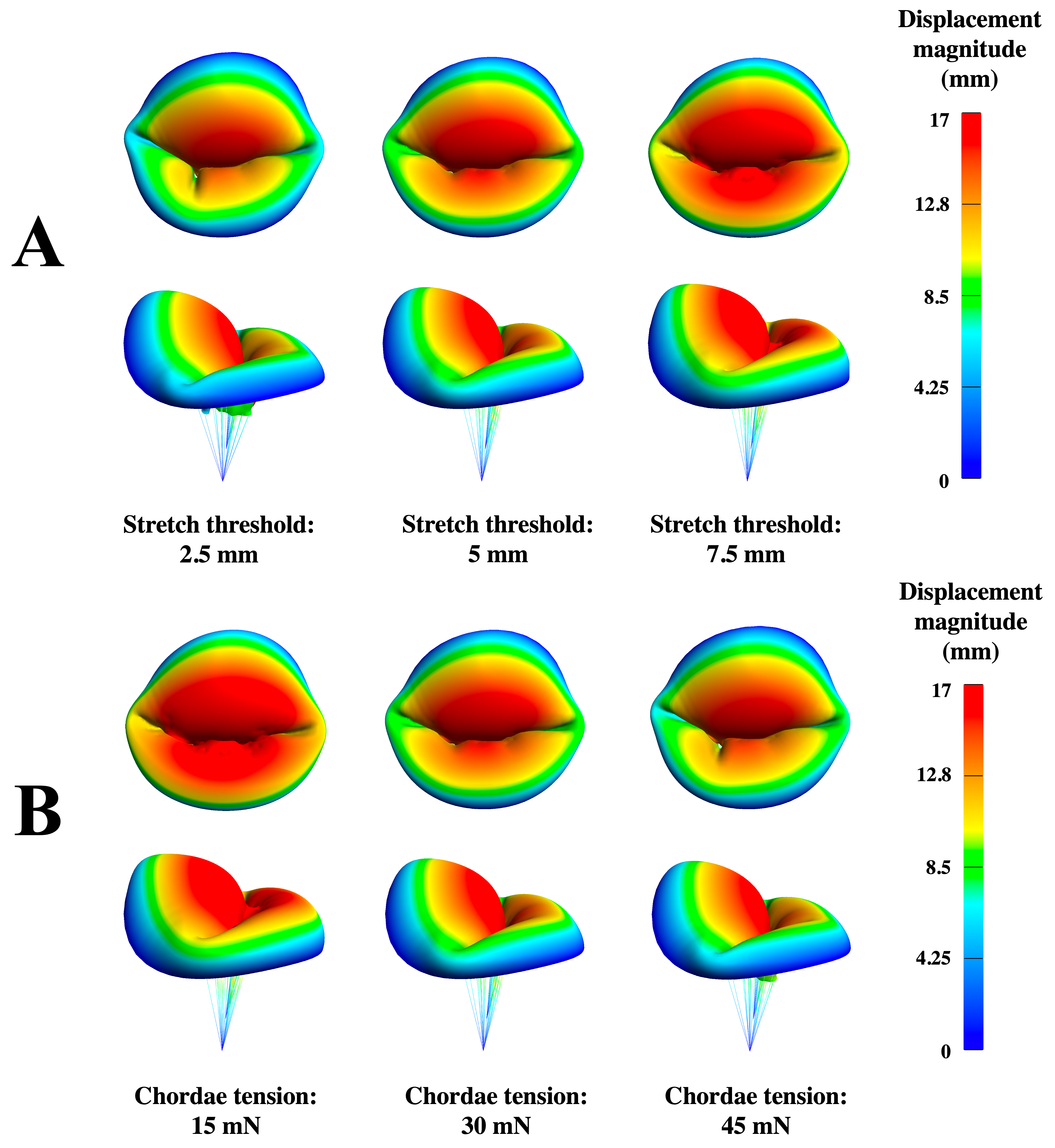}
\caption{Mitral valve systolic configurations resulting from variations in chordae modeling parameters. (A) Systolic configurations subject to various stretch threshold with a fixed chordal tension at 30 mN; and (B) systolic configurations subject to various chordal tension with a fixed stretch threshold at 5 mm. Higher stretch threshold or lower chordal tension force led to models with noticeable billowing.}
\label{mitral_chordae}
\end{figure}

\begin{figure}[h!]
\centering
\includegraphics[width=0.8\textwidth]{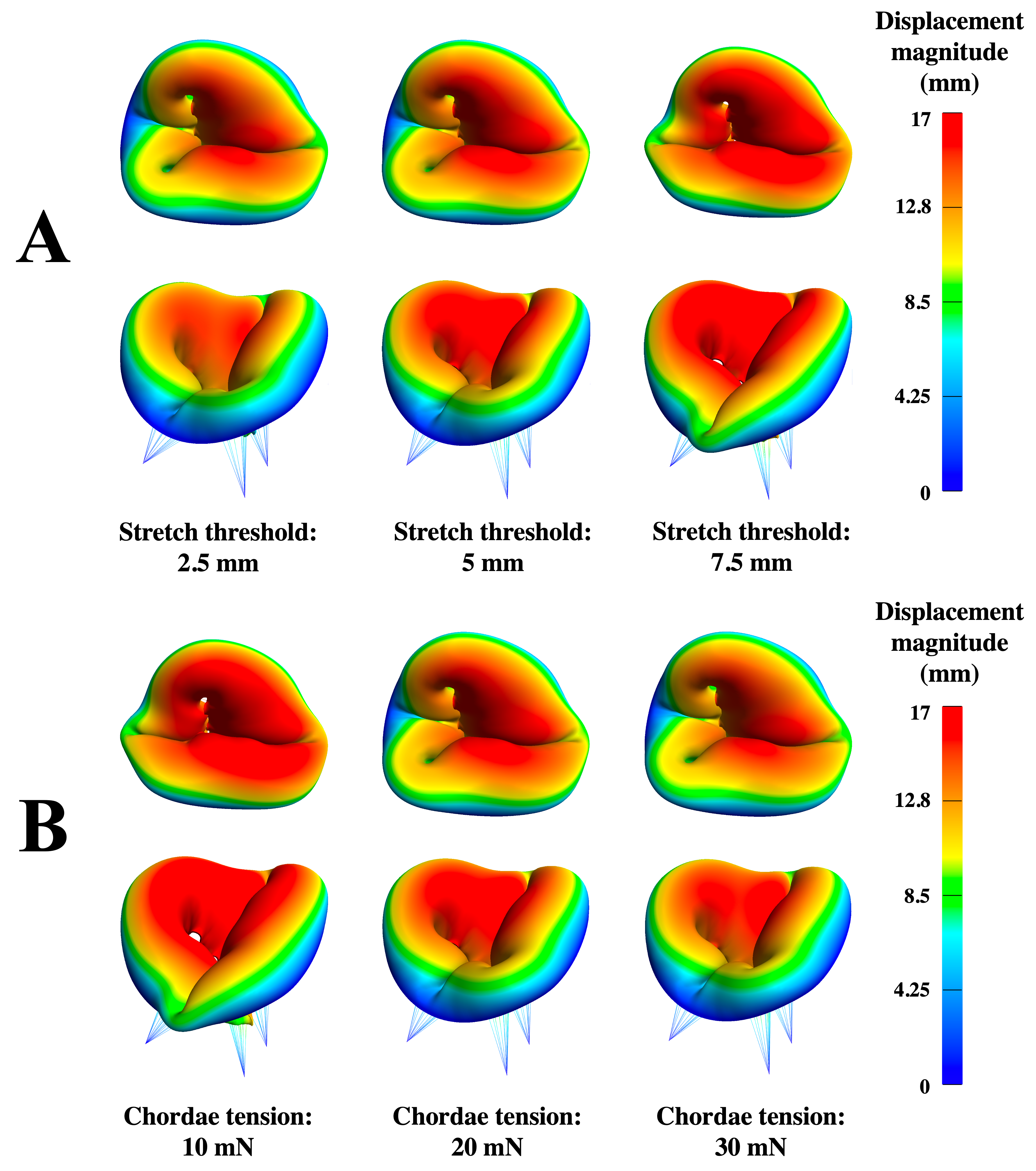}
\caption{Tricuspid valve systolic configurations resulting from variations in chordae modeling parameters. (A) Systolic configurations subject to various stretch threshold with a fixed chordal tension at 20 mN; and (B) systolic configurations subject to various chordal tension with a fixed stretch threshold at 5 mm. Higher stretch threshold or lower chordal tension force led to models with noticeable billowing.}
\label{tricuspid_chordae}
\end{figure}

The systolic configurations of the mitral and tricuspid valves at steady state are presented in Fig. \ref{mitral_chordae} and \ref{tricuspid_chordae} to assess the influence of chordae modeling parameters on valve closure. Models with higher stretch threshold or lower chordal tension force displayed noticeable billowing, or leaflet prolapse into the atrium. Alternatively, models with early or excessive tensioning of the chordae demonstrated poor coaptation, indicating inadequate valve closure, and potential regurgitation.

\newpage
\subsection{Statistical Sensitivity Analysis}
To facilitate the statistical sensitivity analysis, we applied a fourth polynomial order PCE function to identify the sampling points in the material parameter space for FEM analyses. We first assessed the sensitivity indices from interactions between parameters (\textit{i.e.,} the output uncertainty due to variations from two or more parameters). The relative variances obtained were small -- less than 0.005 -- indicating that the material parameters contribute independently to the uncertainty of the model output.

\begin{figure}[h!]
\centering
\includegraphics[width=1\textwidth]{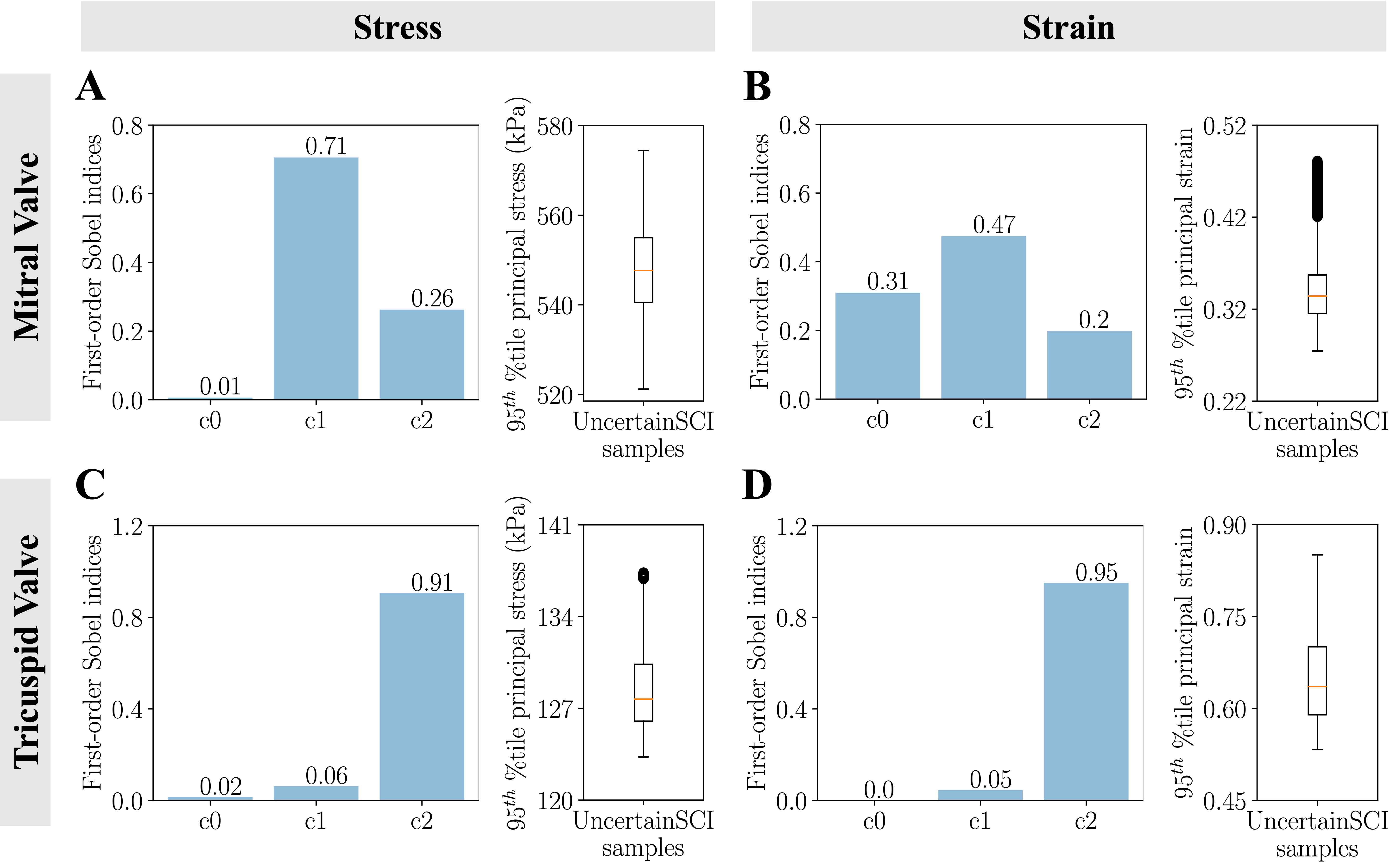}
\caption{The main sensitivity of the stresses and strains as a function of material constants using UncertainSCI. (A) Uncertainty in mitral valve stress, (B) uncertainty in mitral valve strain, (C) uncertainty in tricuspid valve stress, and (D) uncertainty in tricuspid valve strain. Bar plots denote the first-order Sobel indices and box plots depict the spread of stresses and strains of 45 UncertainSCI samples with fixed polynomial chaos expansion order 4 and fixed random number generator seed 0. Results were consistent with the traditional approach, with $c_1$ and $c_2$ influencing mitral valve stresses and strains (with additional influence of $c_0$ on strains), and $c_2$ influencing tricuspid valve stresses and strains.}
\label{SCI_sensitivity}
\end{figure}

The total sensitivities of the material constants were reported in Fig. \ref{SCI_sensitivity}. The statistical results suggested material coefficient $c_1$ was the most dominant, with Sobel index 0.71 for stresses and 0.47 for strains for the mitral valve. Similar to the findings in the traditional approach, material constant $c_0$ had negligible effects on mitral valve stresses (Fig. \ref{SCI_sensitivity}A), but it was the second highest contributor to the total output variance on mitral valve strains (Fig. \ref{SCI_sensitivity}B). In the tricuspid model, material coefficient $c_2$ contributed the most to the model uncertainty -- $c_2$ had a Sobel index of 0.91 and 0.95 in stresses and strains, respectively. In contrast, the Sobel indices for material coefficients $c_0$ and $c_1$ were well below 0.1 (Fig. \ref{SCI_sensitivity}C and \ref{SCI_sensitivity}D). 

\section{Discussion}
Our long term goal is to develop a robust, open-source computational framework for the modeling of atrioventricular valves from 3D images to inform valve repair in children with congenital heart disease. We have taken the first step to achieve this by integrating and extending established open-source tools for cardiac image processing and finite element analysis \cite{Fedorov2012, Scanlan2018, Nguyen2019, Maas2012, Maas2017, Ateshian2018, Burk2020}. This work forms an initial foundation for the collaborative development and application of biomechanical modeling to investigate the effect of image derived, patient-specific valve structure on leaflet stress and strain. This, in turn, may inform the design and application of more durable valve repairs \cite{Narang2021}.  While our driving application is congenital heart disease, the tools described are fundamentally applicable to any population.

In seminal work on the mitral valve, Votta et al. observed that the stresses on the leaflet belly were in the range of 130 to 540~kPa on the anterior leaflet and 60 to 279~kPa on the posterior leaflet\cite{Votta2008} at the systolic pressure of 120 mm Hg; Wang et al. reported a maximum principal stress of 160~kPa on the anterior leaflet midsection at 110~mm Hg peak systolic pressure \cite{Wang2013}; Lee et al. found the maximum radial and circumferential stresses on anterior belly to be 509.5$\pm$38.4~kPa and 301.4$\pm$12.2~kPa at 90~mm Hg peak transvalvular pressure\cite{Lee2014}. In our study, the 95$^\textrm{th}$ percentile principal stresses on the anterior and posterior belly at 100~mm Hg peak pressure were 536.16~kPa and 300.47~kPa, respectively. Our results agreed well with those reported in the literature;  the differences in the stress values can be explained by differences in image-derived model structure and modeling approaches, including  different material constitutive models, transvalvular pressure, and chordal tension force. As demonstrated in the sensitivity analysis, variations in chordal tension force and stretch threshold can lead to more than 50~kPa difference in principal stresses. In addition, our analyses suggested that overall, the anterior leaflet had higher principal stresses than the posterior leaflet, consistent with prior studies.

Of the previous studies on the FEM modeling of tricuspid valves, Stevanella et al. reported higher maximum principal anterior leaflet stress (430~kPa) than posterior leaflet (120~kPa) at 23.7~mm Hg \cite{Stevanella2010}; Kong et al. reported average principal stress of 37-80~kPa, 25-91~kPa, 24-63~kPa on the anterior, posterior, and septal leaflets, respectively, at mid-systole under the same peak transvalvular pressure \cite{Kong2018}; Laurence et al. reported the von Mises stress of 24.7$\pm$7.9~kPa (anterior), 30.6$\pm$10.9~kPa (posterior), and 41.9$\pm$8.6~kPa (septal) on the tricuspid leaflet belly at 25~mm Hg peak pressure \cite{Laurence2020}. Our tricuspid stress values agreed well with Kong et al.'s approximations. The discrepancies between our results and Stevanella et al. may be due to differences both the geometry of the model and the material properties of their model. (Stevanella et al. applied mitral valve material properties for the tricuspid FEM models.) Meanwhile, Laurence et al.'s tricuspid model was approximately a third smaller than ours, which contributed to the slightly lower stresses reported in their work. In addition to the stresses, Stevanella et al. reported circumferential and radial strains of 0.13-0.16 and 0.25-0.30 on the anterior leaflet belly \cite{Stevanella2010}; Kong et al. reported average principal strains of 0.19-0.26, 0.07-0.17, and 0.11-0.21 on the anterior, posterior, and septal leaflets, respectively \cite{Kong2018}; and Laurence et al. reported maximum principal strains of 0.33$\pm$0.07 (anterior), 0.41$\pm$0.06 (posterior), and 0.44$\pm$0.03 (septal) on the tricuspid leaflet belly. Our strain predictions were slightly higher than the reported values. Given that our sensitivity analysis suggested the chordal stretch threshold and tension force were inconsequential to tricuspid valve strains, the differences observed could be due a combination of differences in constitutive models and valve geometry. 

The valve geometries in the present work were derived from images but simplified to create representative models for this initial investigation. However, anatomically accurate valve geometries are critical in determining the stress profiles on the mitral leaflets \cite{Khalighi2017, Sacks2019, Laurence2020}.  For example,  Jimenez et al. suggested that the annular geometry had direct consequence to the chordal force distribution \cite{Jimenez2005}. Sacks et al. \cite{Sacks2019} observed that the segment A2 had the highest stresses on some models. While we also observed high stress concentration on segment A2, in our specific mitral model segments A1 and A3 had slightly higher stresses -- possibly due to variations in the annular and leaflet geometry in our specific model. Notably, the stress distribution of our mitral valve FEM model agreed well with the model in a previous study with similar mitral annular shape \cite{Prot2009}. This highlights the importance of image-derived patient-specific FEM models in order to generate insights relevant to that specific valve geometry. 

Computational modeling of atrioventricular valve dynamics has historically been a challenge due to numerical instability that arises from leaflet contact. Frequently, researchers had to sacrifice solution accuracy (as obtaining a converged solution was difficult) by terminating simulations based on a fixed number of iterations within each time step rather than by a set residual \cite{Kamensky2015, Aggarwal2016}. Although highly refined commercial packages such as LS-DYNA \footnote{http://www.lstc.com/products/ls-dyna} and ABAQUS \footnote{https://www.3ds.com/products-services/simulia/products/abaqus} made obtaining converged solutions possible, any such simulations required a time step of $10^{-6}$~s or less and took more than 20 days to run with explicit time integration scheme \cite{Morganti2015}. Kamensky's volume potential approach \cite{Kamensky2018} introduced a novel modeling method in contact mechanics, providing a robust and computationally efficient strategy to overcome the challenges encountered in atrioventricular valve FEM modeling. We have now implemented this powerful methodology in FEBio. Together with the implicit time integration scheme available in FEBio, our models were able to achieve satisfactory systolic configurations in well under an hour using a single CPU with 24 cores.

Computational models of atrioventricular valves have become increasingly accurate at representing physical reality. However, these simulations often do not capture the impact of parameter uncertainty in their predictions \cite{Rupp2020}. As such, we performed uncertainty analysis using both traditional and statistical approach to determine the most significant tissue material constants and chordal configuration on the biomechanical response of our mitral and tricuspid valve FEM models. Our results indicated that the material constants in the Lee-Sacks model, as well as the chordal stretch threshold and tension, may have varying degrees of influence in image-derived models of the tricuspid and mitral valve. This highlights 1) the need for population specific, and possibly patient-specific tissue constitutive models for atrioventricular valves, and 2) the importance of anatomically accurate chordal length and properties for the most accurate assessment of the leaflet stress and strain. 

 Consistent with our analysis, prior work has demonstrated that valve dynamics are highly influenced by chordal length and stretch threshold \cite{Mansi2012, Grbic2017}. As such, substantial effort has been dedicated to accurately reconstructing the geometry and topology of chordae tendineae using high-resolution micro-CT imaging of static excised animal hearts \cite{Wang2013, Lee2015, Sacks2019}. Notably, micro-CT cannot currently be applied to living humans or beating hearts. Further, it is not currently feasible to visualize individual chordae in living humans using readily available 4D imaging techniques such as 3DE. As such, several approaches have used to approximate chordal length and geometry in the absence of a detailed knowledge of chordal structure. Mansi et al. approximated the chordal length using the distance between papillary muscle tips and leaflet free edge in end diastole \cite{Mansi2012}. Kong et al. first assumed the chordae were straight and branch-less, then iteratively adjusted the chordal length until the FEM leaflet model matched CT images of the leaflets in systole. \cite{Kong2018}. Khalighi et al. developed a simplified, but functionally equivalent, framework for modeling the chordae tendineae when they cannot be visualized in the 3D image \cite{Khalighi2019}. Specifically, Khalighi et al. proposed to approximate the mitral chordae tendineae topology and geometry as branch-less chords uniformly distributed on the leaflets. The chordal lengths were estimated as the distance from the papillary muscle tips to the leaflet insertion points at systolic configuration. While this method provided accurate leaflet stress and strain response for mitral valve, validation in application to the tricuspid valve has yet be demonstrated. In our work, we adopted a similar approach as Khalighi et al. to model the topology of the chordae tendineae. The applied chordal tension force was obtained through an iterative process until the total tethering force on the papillary muscle tips agreed with those reported in the literature and created realistic valve closure in comparison to the images from which they were derived. The total tethering forces in our mitral valve and tricuspid valve FEM models were 6.84~N and 2.38~N, respectively. These values fall within the range of total tethering forces in the literature, which were 4 to 13.5~N for the mitral valve \cite{Sacks2019, Pham2017, Votta2008, Stevanella2010}, and 2.02 to 4.95~N for the tricuspid valve \cite{Kong2018}. 

Finally, we demonstrated the integration of a recently developed statistical uncertainty analysis toolbox, UncertainSCI, into FEBio. Our UnscertainSCI-based statistical analysis agreed well with the findings from traditional analysis, which demonstrated the fidelity of the WAFP-based PCE method for uncertainty quantification. This WAFP-based PCE function was able to efficiently quantify the sensitivity of model input parameters to model response with significantly fewer sample points than standard Monte Carlo methods \cite{Burk2020}. In addition, this approach offered further insights into the exact uncertainty measures from each input parameters, which were not quantifiable using the traditional approach. This provided additional information regarding the sensitivity that each parameter induced in the model. We computed the main sensitivity analysis with various PCE orders and random number generators (Appendix \ref{appendix:b}). We did not observe significant differences in the sensitivity indices among the PCE orders and random number generators. This indicates the reliability of the WAFP-based PCE method for predicting sensitivities in our present work and in future applications.

\section{Limitations and Future Work}
Valve geometry contributes greatly to the biomechanical function of the valve. In the future, we hope to perform more clinically relevant studies, including comparison of dysfunctional valves to normal valves. In addition, the alteration of valve structure can be used to understand the effect of valve geometry on valve stress and strain. Finally, ``surgical" alteration of such valves, emulating existing and novel repair techniques could be used to optimize and inform surgical repairs before they are attempted in humans.

To the authors' knowledge, there is currently no open-source, ready-to-use mesh generation package for lofting valve surfaces from contour curves. While the main components of model creation are open-source, we utilized a commercial CAD program to mesh the leaflets. In the future, we hope to expand the capability to create and alter image-derived valve shell models directly within SlicerHeart \cite{Kong2018_1, Johnson2021}.

Finally, we utilized constitutive models derived from adult mitral and tricuspid valves. While this allows realistic comparison to existing work, and a means for comparison of the valve biomechanics with variation of a baseline structure, it is likely that the adult-derived constitutive models do not accurately describe the mechanical properties of the wide range of ages and valve types present in congenital heart disease. Further work is needed to develop a framework for the development of age and pathology-specific constitutive models for this diverse population, as well as refinement of applicable leaflet shape fitting approaches to mitigate this impediment to translational application when precise constitutive models are not available \cite{Narang2021}.

\section{Conclusion} 

We describe the preliminary implementation of an integrated image-to-FEM-modeling workflow for the biomechanical modeling of atrioventricular valves. While the driving application underlying the development of this evolving open-source framework was to inform a more disciplined and rigorous approach to the assessment of valve failure in children congenital heart disease, it is fundamentally applicable to a wide range of valve science. Our initial stress and strain results yielded excellent agreement compared with the literature and we provided a detailed sensitivity analysis of the FEM modeling parameters using both traditional and statistical methods. Future work will focus on optimization, validation, and application to investigate the biomechanics of dysfunctional atrioventricular valves. 

\section{Acknowledgment}
This work was supported by NIH R01HL153166, R01GM083925, U24EB029007, Big Hearts to Little Hearts, a Children's Hospital of Philadelphia (CHOP) Frontier Fund (Pediatric Valve Center), The Cora Topolewski Fund at CHOP Pediatric Valve Center, and the Canarie Research Software foundation.

\bibliographystyle{unsrt}  
\bibliography{library}

\newpage
\appendix
\section{Appendix A: Additional Verification Results} \label{appendix:a}
Here, we report on additional stress and strain data for the mitral and tricuspid FEM models.

\begin{table}[h!]
\centering
\resizebox{\textwidth}{!}{\begin{tabular}{ccccccc}
\toprule
 && \multicolumn{2}{c}{$1^\textrm{st}$ principal stress} & \multicolumn{3}{c}{$1^\textrm{st}$ principal strain} \\\cline{2-7}
      & mean     & $75^\textrm{th}$ percentile & $95^\textrm{th}$ percentile & mean      & $75^\textrm{th}$ percentile & $95^\textrm{th}$ percentile\\
\midrule  
Mitral  &  252.29   &  336.45  &  546.22   &    0.21      &     0.25  &     0.34  \\
\rowcolor{lightgray}
A1    &  489.89     &  572.02    &  635.20  &     0.31     &     0.35     &     0.38  \\
A2   &  474.85    &  504.18     &  536.16   &     0.27     &     0.29 &     0.32     \\
\rowcolor{lightgray}
A3  &  522.06      &  566.63    &  618.20   & 0.31  &     0.34    &     0.37 \\
P1   &  244.96 &  260.34  &  276.73 &   0.22     &    0.23   &   0.25 \\
\rowcolor{lightgray}
P2     & 267.06     &  282.82   &  300.47  &     0.23    &    0.24 &    0.24     \\
P3  &  248.22   &  264.90 &  276.13  &    0.21    &     0.23   & 0.25  \\
\bottomrule
\end{tabular}}
\caption{$1^\textrm{st}$ principal stresses and strains on the mitral valve with the fine mesh. We observed the highest $95^\textrm{th}$ percentile 1$\textrm{st}$ principal stress and strain in region A1.} \label{mitral_responses_summary} 
\end{table} 

\begin{table}[h!]
\centering
\resizebox{\textwidth}{!}{\begin{tabular}{ccccccc}
\toprule
 && \multicolumn{2}{c}{$1^\textrm{st}$  principal stress} & \multicolumn{3}{c}{$1^\textrm{st}$  principal strain} \\\cline{2-7}
      & mean     & $75^\textrm{th}$ percentile & $95^\textrm{th}$ percentile & mean      & $75^\textrm{th}$ percentile & $95^\textrm{th}$ percentile\\
\midrule  
Tricuspid  &  64.83   &  82.62 &  127.27   &    0.51     &     0.59  &     0.63  \\
\rowcolor{lightgray}
Anterior    &  75.53     &  101.44    &  142.88 &     0.52    &     0.61     &     0.64 \\
Posterior   &  68.76    &  86.71    &  112.94  &     0.51    &     0.58 &     0.62    \\
\rowcolor{lightgray}
Septal  &  51.48     &  65.08   &  90.20   & 0.51  &     0.58    &    0.61 \\
\bottomrule
\end{tabular}}
\caption{$1^\textrm{st}$  principal stresses and strains on the tricuspid valve  with the fine mesh. We observed the highest $95^\textrm{th}$ percentile 1$\textrm{st}$ principal stress and strain in the anterior leaflet.} \label{tricuspid_responses_summary} 
\end{table} 

\newpage
\section{Appendix B: FEBioUncertainSCI Sensitivity} \label{appendix:b}
Here, we report on the sensitivity of FEBioUncertainSCI to the polynomial order and random number generator.

\begin{figure}[h!]
\centering
\includegraphics[width=1\textwidth]{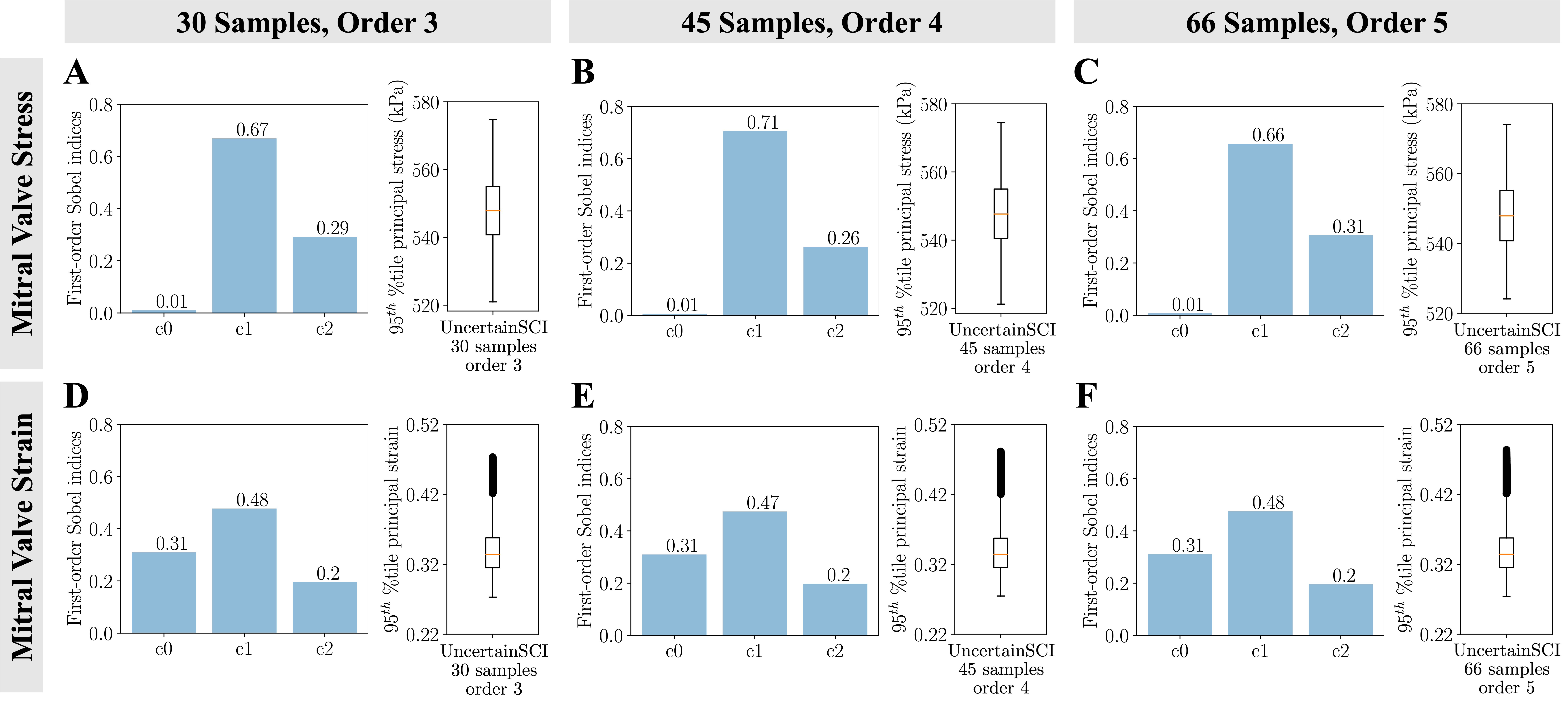}
\caption{The main sensitivity of the mitral valve stresses and strains as a function of material constants with varying polynomial chaos expansion (PCE) orders using UncertainSCI. (A) Uncertainty in valve stress with 30 samples and PCE order 3, (B) uncertainty in valve stress with 45 samples and PCE order 4, (C) uncertainty in valve stress with 66 samples and PCE order 5, (D) uncertainty in valve strain with 30 samples and PCE order 3, (E) uncertainty in valve strain with 45 samples and PCE order 4, and (F) uncertainty in valve strain with 66 samples and PCE order 5. Bar plots denote the first-order Sobel indices and box plots depict the spread of stresses and strains of UncertainSCI samples with a fixed random seed of 0. We did not observe significant differences in the first-order Sobel indices.}
\label{mitral_SCI_sensitivity_pce}
\end{figure}

\begin{figure}[h!]
\centering
\includegraphics[width=1\textwidth]{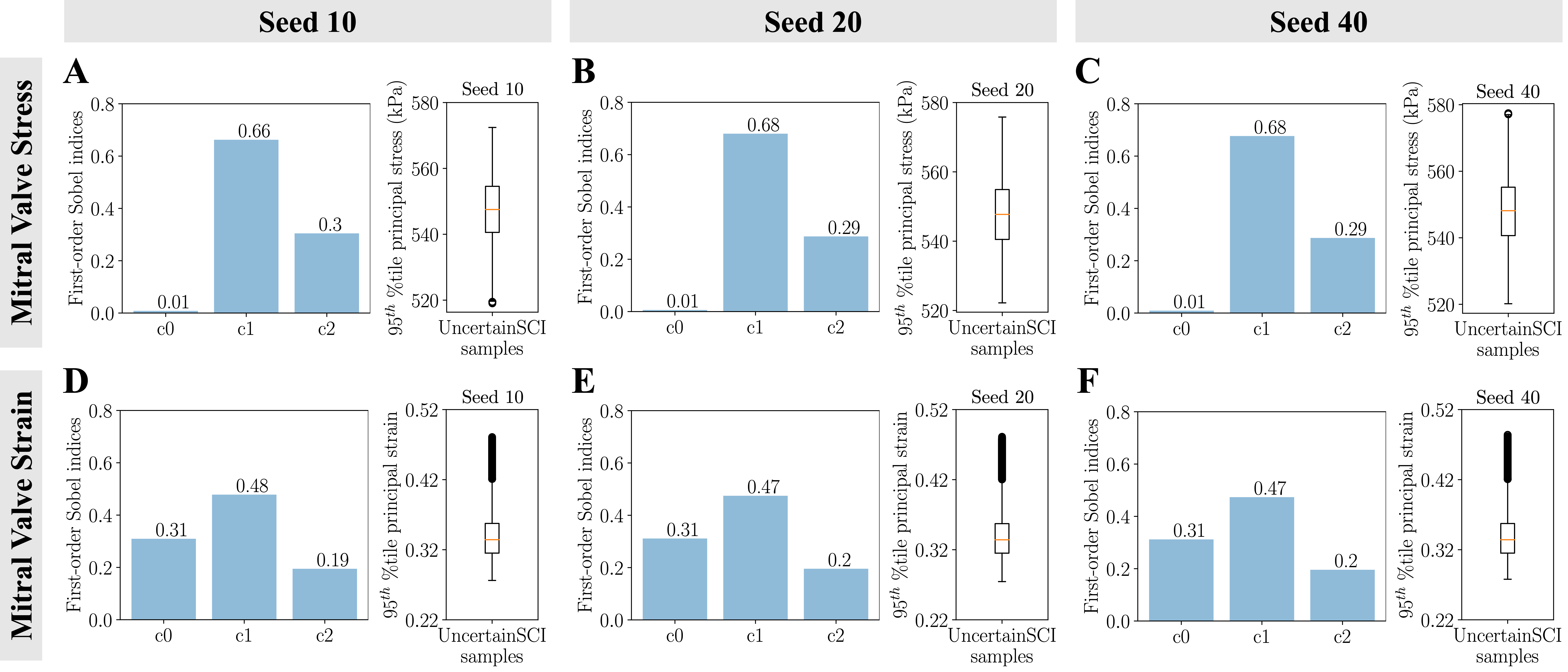}
\caption{The main sensitivity of the mitral valve stresses and strains as a function of material constants with varying random number generator seeds using UncertainSCI. (A) Uncertainty in valve stress with random seed 10, (B) uncertainty in valve stress with random seed 20, (C) uncertainty in valve stress with random seed 40, (D) uncertainty in valve strain with random seed 10, (E) uncertainty in valve strain with random seed 20, and (F) uncertainty in valve strain with random seed 40. Bar plots denote the first-order Sobel indices and box plots depict the spread of stresses and strains of 45 UncertainSCI samples with fixed polynomial chaos expansion order 4. We did not observe significant differences in the first-order Sobel indices.}
\label{mitral_SCI_sensitivity_seed}
\end{figure}

\begin{figure}[h!]
\centering
\includegraphics[width=1\textwidth]{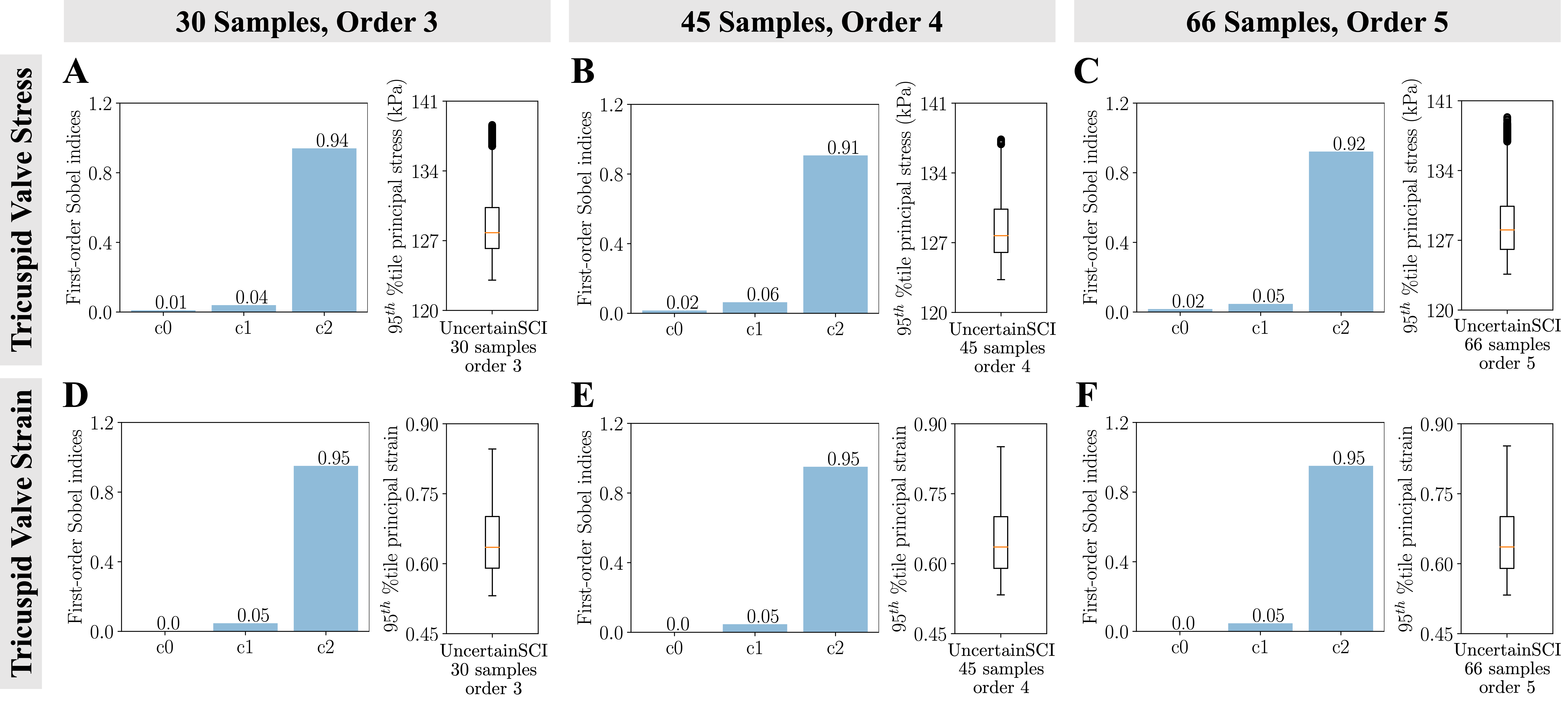}
\caption{The main sensitivity of the tricuspid valve stresses and strains as a function of material constants with varying polynomial chaos expansion (PCE) orders using UncertainSCI. (A) Uncertainty in valve stress with 30 samples and PCE order 3, (B) uncertainty in valve stress with 45 samples and PCE order 4, (C) uncertainty in valve stress with 66 samples and PCE order 5, (D) uncertainty in valve strain with 30 samples and PCE order 3, (E) uncertainty in valve strain with 45 samples and PCE order 4, and (F) uncertainty in valve strain with 66 samples and PCE order 5. Bar plots denote the first-order Sobel indices and box plots depict the spread of stresses and strains of UncertainSCI samples with a fixed random seed of 0. We did not observe significant differences in the first-order Sobel indices.}
\label{tricuspid_SCI_sensitivity_pce}
\end{figure}

\begin{figure}[h!]
\centering
\includegraphics[width=1\textwidth]{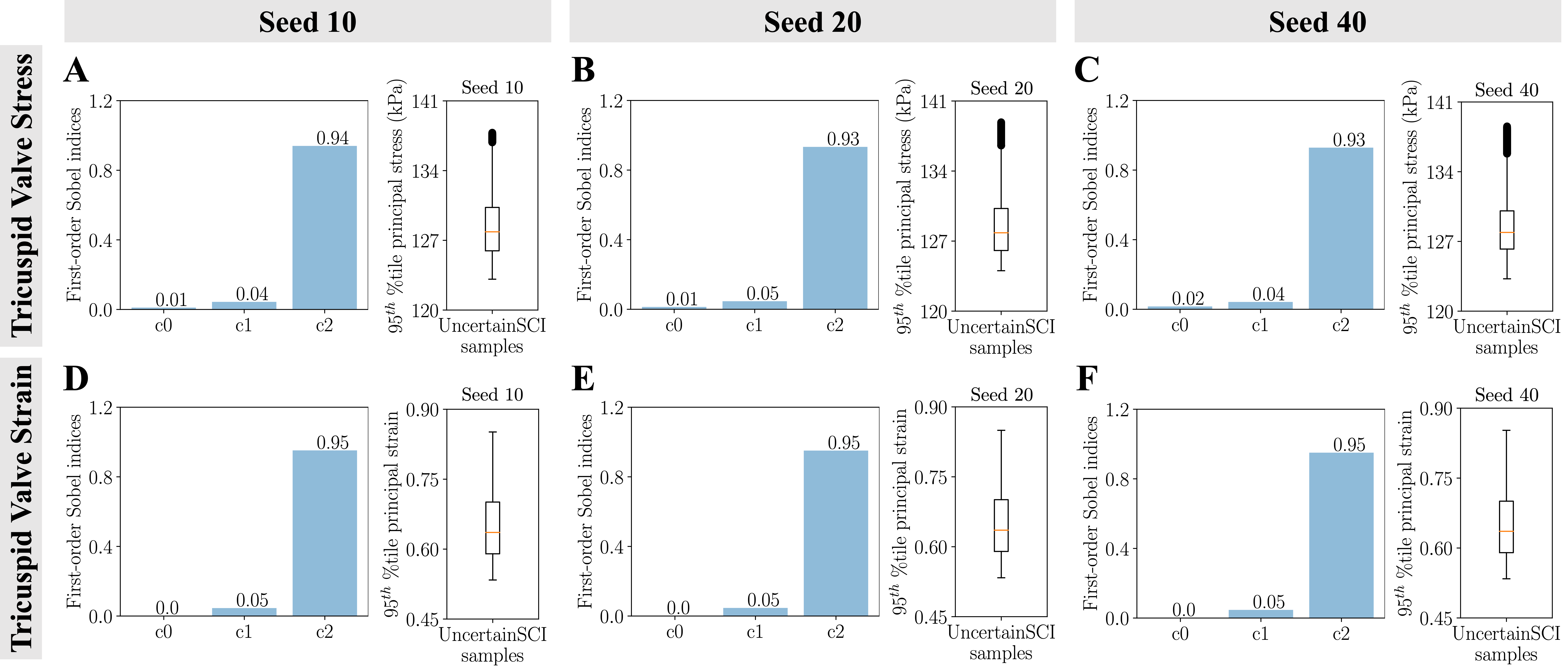}
\caption{The main sensitivity of the tricuspid valve stresses and strains as a function of material constants with varying random number generator seeds using UncertainSCI. (A) Uncertainty in valve stress with random seed 10, (B) uncertainty in valve stress with random seed 20, (C) uncertainty in valve stress with random seed 40, (D) uncertainty in valve strain with random seed 10, (E) uncertainty in valve strain with random seed 20, and (F) uncertainty in valve strain with random seed 40. Bar plots denote the first-order Sobel indices and box plots depict the spread of stresses and strains of 45 UncertainSCI samples with fixed polynomial chaos expansion order 4. We did not observe significant differences in the first-order Sobel indices.}
\label{tricuspid_SCI_sensitivity_seed}
\end{figure}

\end{document}